\documentclass[showpacs,preprintnumbers,amsmath,amssymb]{revtex4}

\usepackage{graphicx}
\usepackage{dcolumn} 
\usepackage{bm}      
 
\newcommand{\e}[1]{$\times10^{#1}$}

\begin{document}

\title{ Small localized  black holes in a braneworld: 
\\Formulation and numerical method
}

\author{Hideaki Kudoh${^{1}}$ }
\email{kudoh@yukawa.kyoto-u.ac.jp}
\author{Takahiro Tanaka${^{2}}$} 
\email{tanaka@yukawa.kyoto-u.ac.jp}
\author{Takashi Nakamura${^{1}}$}
\email{takashi@yukawa.kyoto-u.ac.jp}

\affiliation{  ${^{1}}$  Department of Physics, Kyoto University, Kyoto 606-8502, Japan 
\\
 ${^{2}}$ Yukawa Institute for Theoretical Physics, 
             Kyoto University, Kyoto 606-8502, Japan}

\begin{abstract}
No realistic black holes localized on a 3-brane in the Randall-Sundrum infinite braneworld have been found so far. 
The problem of finding a static black hole solution is reduced to a boundary value problem. We solve it by means of a numerical method, and show numerical examples of a localized black hole whose horizon radius is small compared to the bulk curvature scale. The sequence of small localized black holes exhibits a smooth transition from a five-dimensional Schwarzschild black hole, which is a solution in the limit of small horizon radius. 
The localized black hole tends to flatten as its horizon radius increases. 
However, it becomes difficult to find black hole solutions as its horizon radius increases.
\end{abstract}
\pacs{04.50.+h, 04.70.Bw, 04.70.Dy}

\preprint{KUNS-1815, YITP-03-3}

\maketitle


\section{Introduction}
Higher-dimensional black holes have been considered for a long time as purely theoretical applications motivated by higher-dimensional theories, such as string theory. However, recent developments in the scenario of large extra dimensions ~\cite{Arkani-Hamed:1998rs} have aroused new interest in such black holes. In the braneworld scenario, an interesting possibility of black hole production at a collider was pointed out
\cite{Giddings:2001bu,Dimopoulos:2001hw} (see, e.g., \cite{Cavaglia:2002si} and references therein). There are other types of braneworld models proposed by Randall and Sundrum (RS) \cite{Randall:1999ee,Randall:1999vf}. 
In these models, the geometry warped in the direction of an extra dimension is used to explain the hierarchy between the TeV scale and the Planck scale, and to realize four-dimensional gravity effectively on the 3-brane. 
Also in the context of RS models, higher-dimensional black holes may play an important roll. 
These so-called braneworld scenarios provide new and interesting situations to investigate higher-dimensional black holes.

In the model of large extra dimensions, a physically meaningful sequence of black hole solutions will be obtained as a slight modification of the higher-dimensional Kerr black hole  \cite{Myers:un} (or more simply the Schwarzschild black hole  \cite{Tangherlini,Myers:rx,Bogojevic:1990hv})  if the horizon radius is sufficiently small compared to the extension of extra dimensions and the self-gravity due to periodic boundary is weak. 
As another sequence, there are the black string solutions. 
A black string is in general unstable to linear perturbations with long wavelength in the direction along the string, which is called Gregory-Laflamme instability~\cite{Gregory:1993vy}. 
(A stability analysis of black strings is also found in Ref.~\cite{Hirayama:2001bi}.) 
Therefore, a black string is unstable when the horizon radius of the black string is sufficiently small compared to the extension of the extra dimension. 
For such a small black hole, the former sequence is expected to be stable. 
For RS models, since the 3-brane has tension, it is more difficult to find black hole solutions. 
Trivial black string solution is allowed also in these models, and it becomes unstable in the same way. 
There are many discussions about black holes in this model and some black hole solutions have been considered by several authors~\cite{Dadhich:2000am,Giannakis:2001ss,Chamblin:2001ra,Cadeau:2001tj,Vacaru:2001rf,Vacaru:2001wc,Kanti:2001cj,Casadio:2001jg,Casadio:2002uv,Sengupta:2002fr,Modgil:2001hm,Kofinas:2002gq,Kodama:2002kj,Chamblin:2000by,Emparan:2001wk,Emparan:2001wn}.
A strategy to construct a black hole solution is to assume an induced metric on the 3-brane as {\it initial data}, and extend it to the bulk analytically or numerically.  
This method generally results in a naked singularity in the bulk since there is no guarantee that the induced metric assumed as a boundary condition is compatible with a regular geometry. 
If we randomly specify the boundary metric on the 3-brane, almost all solutions develop a naked singularity. 
After all, no realistic black hole solutions which are stable and have no naked singularity have been found so far, and finding them is an interesting open question of nonlinear gravity in the braneworld.

Successful recovery mechanism of 4D gravity on the 3-brane \cite{Shiromizu:2000wj,Garriga:2000yh,Tanaka:2000er,Mukohyama:2001ks,Tanaka:2000zv,Giannakis:2001zx,Kudoh:2001wb,Kudoh:2001kz,Kudoh:2002mn} suggests the existence of black holes in the RS infinite braneworld~\cite{Randall:1999vf}.
The shape of black holes is conjectured based on the
Gregory-Laflamme instability~\cite{Chamblin:2000by}. It was argued that an unstable black string will be pinched into many black holes. 
In addition to black holes in the bulk, there will be a black hole that is localized on the 3-brane. We can imagine this as a black hole bound to a domain wall if the 3-brane is realized by a domain wall~\cite{Emparan:2000fn}. 
An exact solution representing a localized black hole is known in the 4D braneworld model, which we call in this paper the Emparan-Horowitz-Myers (EHM) solution~\cite{Emparan:2000wa}.
However the corresponding solution in the original 5D braneworld model has not been discovered. 
While the localization of a black hole was motivated by the classical instability of a black string, the anticipated dynamics of pinching off the horizon is questioned by Horowitz and Maeda~\cite{Horowitz:2001cz}. 
Recently, Wiseman discovered nonuniform black strings by numerical calculation~\cite{Wiseman:2002zc}. 
The obtained solutions are likely to be unstable, but they are suggestive of a missing link between the localized black hole and the black string. 
On the other hand, based on an extensive use of the AdS/CFT correspondence, there is another discussion anticipating the absence of static localized black holes~\cite{Tanaka:2002rb,Emparan:2002px}.

The present paper explores the problem of black holes in the RS infinite braneworld. 
We consider a numerical construction of a black hole solution in this model.  
The method we use in this paper is based on a scheme developed by Wiseman~\cite{Wiseman:2001xt}.  
Our method does not require any assumptions for the induced metric on the 3-brane. 
We solve Einstein equations numerically under the boundary conditions determined by physical requirements. 
We will find small black hole solutions, whose horizon radius is smaller than the AdS curvature radius, although large black hole solutions have not been discovered.

In the next section, we explain our method formulating the problem to be suitable for numerical calculation, and the boundary conditions are also discussed. The examples of numerical solutions are shown in Sec. \ref{sec:3}. 
Section \ref{sec:Discussion} is devoted to a summary.

\section{ Formulation as a boundary value problem }

\subsection{Conformal transformation}

To obtain black hole solutions in the RS infinite braneworld model, it is crucial to formulate the problem as a boundary value problem without assuming any artificial boundary conditions.  We follow and develop the numerical method developed in Ref.~\cite{Wiseman:2001xt} (see also~\cite{Wiseman:2002zc}). 
What we want to find is a static black hole solution that is localized on the brane. For simplicity, we further restrict our attention to the configuration with 5D axial symmetry (4D spherical symmetry). 
Under the static and axisymmetric assumption, the metric depends only on the radial coordinate $r$ and the coordinate $z$ in the direction of the extra dimension. 
Then, without loss of generality, the metric can be written as 
\begin{eqnarray}
ds^2 = \frac{\ell^2}{z^2} 
\left[
  - T^{2} dt^2 +e^{2R } (dr^2+dz^2)+r^2 e^{2C} d\Omega ^2
\right] \,,
\label{eq:assume}
\end{eqnarray}
where 
$d\Omega ^2=d\theta^2+\sin^2 \theta d\phi^2$ represents the line element on a unit 2-sphere.
The cosmological constant in the bulk is related to the bulk curvature length $\ell$ by $\Lambda= -6/\ell^2$. 
If we set $T=1$ and $R=C =0$, this metric becomes the AdS metric in the Poincar\'e coordinates. 
Since we consider a localized black hole, polar coordinates 
\begin{eqnarray}
&& r=\rho \sin\chi,\cr
&& z=\ell+\rho\cos\chi,\cr
&& \xi=\chi^2,
\end{eqnarray}
are more convenient. 
Then, we have 
\begin{eqnarray}
&& dr^2+dz^2 
 = d\rho^2 + \frac{\rho^2}{4\xi}d\xi^2.
\end{eqnarray}
The angular coordinate $\xi$ is useful to treat the coordinate singularity at $\chi=0$ numerically since the singularity becomes milder in $\xi$ coordinate as we will see later~\cite{Nakamura:1981kc}.

For numerical calculations, it is convenient if the boundaries are located on lines where one of the coordinates is constant.
This can be achieved in general by using the residual gauge degrees of freedom.
The metric form (\ref{eq:assume}) has the gauge degrees of freedom of conformal transformations in the two-dimensional space spanned by  
$r$ and $z$.  
For our present purpose, it is convenient to use the conformal polar coordinates $(\zeta,\chi)$, where $\zeta=\log \rho$. Using these coordinates, we have $dr^2+dz^2=\rho^2(d\zeta^2+d\chi^2)$. 
For the conformal transformation 
$\zeta' =f(\zeta,\chi)$ and $\chi'= g(\zeta,\chi)$,  
$f$ and $g$ satisfy the Cauchy-Riemann relations. 
With a function $\psi(\zeta,\chi)$, we have 
\begin{eqnarray}
 {d\zeta'}^2 +{d\chi'}^2 =  \psi (\zeta,\chi) \left(d\zeta^2 + d\chi^2\right). 
\end{eqnarray}
We can use these conformal degrees of freedom to deform the shape of the boundaries. 
From the Cauchy-Riemann relations, $g$ (and also $f$) satisfies Laplace equation, and thus $g$ is completely determined by specifying its boundary conditions. Suppose we set $g=0$ on the axis and $g=\pi/2$ on the brane, and impose Neumann boundary conditions on the horizon boundary and the asymptotic boundary.  The transformation $f$ is determined by integrating the Cauchy-Riemann relations. 
The Neumann conditions for $g$ on the horizon boundary and on the asymptotic boundary guarantee the constancy of $f$ on these two boundaries. 
In this manner, we can generally find a conformal transformation that transforms each boundary to a constant coordinate line.  
Therefore, the location of the event horizon can be transformed to be
\begin{eqnarray}
\rho|_{\mathrm{horizon}} =  \mathrm{const} =  \rho_h \,,
\end{eqnarray}
and the brane can be placed at $z=\ell$~\cite{Garriga:2000yh}.
Here we note that $f$ is uniquely determined up to adding a constant, and hence, the difference of the value of $\zeta'$ between two points cannot be changed arbitrarily. 
We therefore do not have a degree of freedom to change the ratio $\rho/\rho_h$ for a given solution. 
   
\subsection{Elliptic equations and constraint equations}

Let us consider the elliptic equations and constraint equations that we solve.
From the five dimensional vacuum Einstein equations, ${\mathcal G}^\mu_\nu:=R^{\mu}_\nu-\frac{2}{3}\Lambda g_{\mu}^{\nu}=0$, 
 we obtain an elliptic equation for each metric component. 
Respective equations are given from ${\mathcal G}^t_t$, $ 
({\mathcal G}^t_t - {\mathcal G}^\rho_\rho -{\mathcal G}^\xi_\xi
+ 2{\mathcal G}^\theta_\theta)$ and ${\mathcal G}^\theta_\theta$ as 
\begin{eqnarray}
\Delta  T 
&=&
 - \frac{4 T}{z^2} \left(1 + \frac{\Lambda \ell^2}{6} e^{2R} \right) 
 - \frac{4\chi T_{,\xi}}{\rho^2}  \left(\cot\chi +{2r\over z} +2\chi C_{,\xi}
   \right)
 + 2T_{,\rho}  \left( \frac{1}{\rho}-{2\ell \over \rho z}-C_{,\rho}
   \right)
\nonumber
\\
&& + \frac{2 T}{\rho} \left[ \left(1-\frac{\ell}{z}\right) C_{,\rho} 
     - \frac{2\chi \sin\chi}{z }
   C_{,\xi} \right] \,,
\label{eq:pol-T}
\\
 \Delta R 
&=& 
    \frac{1 - e^{2 (R-C)}}{r^2} 
  +  \frac{2}{z^2} \left(1 + \frac{\Lambda \ell^2}{6} e^{2R} \right) 
  +\frac{4\chi T_{,\xi}}{ \rho^2 T} \left(\cot\chi+{r\over z}+2\chi C_{,\xi}\right) 
  + \frac{2T_{,\rho}}{T}  \left({\ell \over \rho z}+C_{,\rho}\right)
\cr
&&
  + \frac{4\chi C_{,\xi} }{\rho^2} \left(\cot\chi+{2r\over  z} +\chi C_{,\xi}\right)
  + C_{,\rho} \left(- \frac{2}{\rho } +  {4\ell \over \rho z}+ C_{,\rho} \right) \,,  
\label{eq:pol-R}
\\
  \Delta C 
&=& 
  - \frac{1 - e^{2 (R-C) }}{r^2}
  - \frac{4}{z^2} \left(1 + \frac{\Lambda \ell^2}{6} e^{2R} \right) 
  - {2\chi T_{,\xi} \over \rho^2 T}  \left(\cot\chi+{r\over z}+2\chi C_{,\xi}\right)
  - {T_{,\rho}\over T}  \left({\ell\over \rho z}+C_{,\rho}\right)
\cr
&&
  - \frac{2 \chi C_{,\xi}}{\rho^2} \left(4\cot\chi+{5r\over z}+4\chi C_{,\xi}\right)
  + C_{,\rho} \left( \frac{1}{\rho}-{5\ell \over \rho z}-2C_{,\rho}\right)\,,  
\label{eq:pol-C}
\end{eqnarray}
where a comma means a partial derivative, and 
$ 
 \Delta = \partial_r^2 + \partial_z^2 
        =  \partial_\rho^2 + {\rho^{-1}} \partial_\rho +
	2\rho^{-2}(2\xi\partial_\xi^2+ \partial_\xi)
$
is the Laplace operator. 
The first terms in the equations for $R$ and $C$ have the factor that behaves as
$r^{-2} \propto {\chi}^{-2}$ when $\chi \to 0$. Introduction of
$\xi$ changes this severe behavior and it makes the terms more tractable in numerical calculation.
Constraint equations are obtained, respectively, from ${\mathcal G}^\rho_\xi$ and $({\mathcal
G}^t_t - {\mathcal G}^\rho_\rho +{\mathcal G}^\xi_\xi
+ 2{\mathcal G}^\theta_\theta)$ as
\begin{eqnarray}
\Theta_1 &:=&
\xi T \left( \frac{\ell^2 e^{2R}}{z^2}\right)  {\mathcal G}^\rho_\xi\,,
\cr
&=&   \xi \left[
    R_{,\xi}T_{,\rho} - T_{,\xi\rho}+ T_{,\xi} \Bigl(R_{,\rho}+\frac{1}{\rho}\Bigr) 
    \right]
  + T \biggl[ 
  \xi R_{,\xi} \left( 2C_{,\rho} -\frac{1}{\rho} + {3\ell \over \rho z} \right)  
 - 2\xi C_{,\xi\rho}
 -  C_{,\rho} \Bigl(  \frac{\chi}{\tan\chi} + 2\xi C_{,\xi} \Bigr)
  \cr
&&
  + R_{,\rho} \left( \frac{\chi}{\tan\chi} + {3r \sqrt{\xi} \over 2z}+ 2\xi C_{,\xi} \right)
  \biggr] =0 
         \,,
\label{eq:const eq in rz}
\\
\Theta_2 &:=&
- \frac{1}{2} \left( \frac{\ell^2 e^{2R}}{z^2} \right) 
\left( {\mathcal G}^t_t - {\mathcal G}^\rho_\rho + {\mathcal G}^\xi_\xi + 2{\mathcal G}^\theta_\theta \right)
\,,
\cr
&=&     \frac{4\xi T_{,\xi\xi} }{\rho^2} 
  +  \frac{ 2T_{,\xi}}{\rho^2}
    \left( 1 + \frac{2\chi}{\tan\chi} + {3r\sqrt{\xi} \over z} + 2\xi(2 C_{,\xi}-R_{,\xi})\right) 
 + {T_{,\rho}} \left({3\ell\over \rho z}+2C_{,\rho}+R_{,\rho}
         \right)
\cr
&&
 + T \biggl[
  \frac{6}{z^2} \left( 1 - \frac{|\Lambda| \ell^2}{6} e^{2 R} \right)
  + \frac{1 - e^{2 (R-C) }}{r^2 }
  + \frac{8\xi C_{,\xi\xi}}{\rho^2}
  + \frac{12 C_{,\xi}}{\rho^2} \left( {1 \over 3} + \frac{ \chi }{\tan\chi}    
  + {r \sqrt{\xi} \over z}+  \xi C_{,\xi} \right)
\cr 
&&
  + C_{,\rho} \left( {6 \ell \over\rho z}-\frac{2}{\rho} + C_{,\rho} \right)
  - \frac{2 R_{,\xi}}{\rho^2} 
    \left( 
        \frac{2\chi}{\tan \chi} +{3r \sqrt{\xi} \over z} + 4\xi C_{,\xi}
    \right)
  +   R_{,\rho} \left( {3\ell\over \rho z}-\frac{1}{\rho} +2C_{,\rho}\right)
  \biggr] =0 \,.
\label{eq:const eq in zz}
\end{eqnarray}

We perform numerical calculations for the elliptic equations using a relaxation method. The constraint equations are not explicitly solved but they are used as a check of accuracy. Before proceeding further, we discuss an important property of the constraint equations~\cite{Wiseman:2001xt}. 
The constraint equations with an appropriate function multiplied satisfy Cauchy-Riemann relations. They are obtained from the nontrivial components of the Bianchi identities $ \nabla_\mu 
{\mathcal G}^\mu_{~\nu}=0$. Assuming that equations for $T$, $R$, and $C$ are satisfied, i.e., $ {\mathcal G}^t_{t} = {\mathcal G}^\theta_{\theta}
= ({\mathcal G}^\rho_{\rho} + {\mathcal G}^\xi_\xi) =0$, we obtain
Cauchy-Riemann relations for $ {\mathcal G}^\rho_\xi$ and $({\mathcal
G}^\rho_\rho - {\mathcal G}^\xi_\xi)$ as 
\begin{eqnarray}
   \partial_\zeta  {\mathcal U}  
 - \partial_\chi {\mathcal V}   &=&0 \,,
\nonumber
\\
   \partial_\chi  {\mathcal U}  
 + \partial_\zeta {\mathcal V}   &=&0 \,,
\end{eqnarray}
where 
$ \mathcal{ U} := \xi \sqrt{-g}\, {\mathcal G}^\rho_\xi/\sin\theta$,
$\mathcal{ V}  :=  \rho \sqrt{-g} \sqrt{\xi}({\mathcal G}^\rho_\rho 
- {\mathcal G}^\xi_\xi)/(4 \sin\theta) 
$, 
and 
$g:= \mathrm{det} (g_{\mu\nu})$. Then
\begin{eqnarray}
\mathcal{ U} &=& 
\left(\frac{\ell}{z}\right)^3 \frac{\rho^3 e^{2C}(\sin\sqrt{\xi})^2}{2\sqrt{\xi}} ~\Theta_1 \,,
\nonumber
\\
\mathcal{ V} &=&
\left( \frac{\ell}{z}\right)^3 
\frac{\rho^4 e^{2C} (\sin\sqrt{\xi})^2}{4}  ~
\Theta_2 \,.
\end{eqnarray}
Since $\mathcal{ U}$ and $\mathcal{ V}$  satisfy Cauchy-Riemann relations, each of them satisfies the Laplace equation. Hence in principle, if $\Theta_1=0$ is satisfied on all boundaries and provided that $\Theta_2$ vanishes at any one point, the two constraint equations are automatically satisfied in all places as long as the elliptic equations for $T, R$, and $C$ are solved.

\subsection{Boundary conditions} 

In order to solve the elliptic equations derived in the previous section, we must specify the boundary conditions. 
Let us first discuss boundary conditions on the symmetry axis of $r=0$. 
The regularity at $r=0$ of Eqs.~(\ref{eq:pol-R}) and (\ref{eq:pol-C})  requires 
\begin{eqnarray}
R=C  \quad (\mbox{at}~  \xi=0), 
\label{R=C}
\end{eqnarray}
which determines the boundary condition for $R$. 
The axial symmetry requires that the $r$ derivative of the metric functions vanishes at $r=0$. However, as long as we use the $\xi$ coordinate, the finiteness of the derivative in this coordinate automatically guarantees the regularity because $T_{,r} \propto \sqrt{\xi} T_{,\xi}$.  
Hence, we adopt a free boundary condition for $T$ and $C$. 
The values of $T$ and $C$ at $r=0$ are evolved in the same way as the values at an ordinary grid point. The only difference is to use the one-sided differentiation to evaluate the first-order derivatives with respect to $\xi$. The second-order derivatives disappear at $\xi=0$ from the Laplace operator. 
Equation (\ref{eq:const eq in rz}) at $\xi=0$ is trivial from 
Eq.~(\ref{R=C}), while 
Eq.~(\ref{eq:const eq in zz}) reduces to
\begin{eqnarray}
      R_{,\xi}= 3C_{,\xi} +\frac{T_{,\xi} }{T}
     + \rho^2 \left[
         \frac{1-e^{2R}}{z^2} + C_{,\rho}\Bigl( \frac{2\ell-\rho}{2\rho z} + \frac{C_{,\rho}}{2} \Bigr)    
        + \frac{  T_{,\rho}}{2T} 
           \Bigl( \frac{\ell}{\rho z}+ {C_{,\rho}} \Bigr)
         \right]
   \quad (\mathrm{at ~ {\xi=0}} ).
\label{R_{,xi}}
\end{eqnarray}
We can also show that this equation is guaranteed to be satisfied if $R$ and $C$ are solutions of Eqs.~(\ref{eq:pol-R}) and (\ref{eq:pol-C}). 
Although it is not necessary in principle, we use this condition~(\ref{R_{,xi}}) in addition to Eq.~(\ref{R=C}) in order to improve the accuracy of the calculation. We do not solve the evolution equation for $R$ at the ordinary grid points next to the axis in place of imposing this supplementary condition.

On the horizon, the Killing vector $\partial_t$ must become null. 
Therefore, we have 
\begin{eqnarray}
 T = 0  \quad (\mathrm{at}~  \rho=\rho_h) \,,
\label{T=0}
\end{eqnarray}
In the present gauge, the horizon is given by a constant radius, and thus we have 
\begin{eqnarray}
T_{,\xi} = T_{,\xi\xi} =0  \quad (\mbox{at}~  \rho=\rho_h ) \,. \label{T,xi=0}
\end{eqnarray}
Here we assume that metric functions and their derivatives are finite on the horizon. 
With the aid of Eqs.~(\ref{T=0}) and
(\ref{T,xi=0}), the surface gravity on the horizon is given as  
\begin{eqnarray}
 \kappa = e^{-R} T_{,\rho} 
\quad (\mbox{at}~  \rho=\rho_h )\,.
\label{surface grav}
\end{eqnarray}
This condition can be used to impose a Dirichlet boundary condition for $R$ on the horizon. For this purpose, we rewrite it as 
\begin{eqnarray}
R = C|_{\xi=0} + \log \left( \frac{T_{,\rho}}{T_{,\rho}|_{\xi=0}} \right) 
\quad (\mathrm{at~} \rho=\rho_h) \,,
\label{R horizon}
\end{eqnarray}
using Eq.~(\ref{R=C}).

Following the line of a standard proof~\cite{Wald:1984}, we can show  that the zeroth law of black hole thermodynamics, $\partial_\chi \kappa =0$, is valid, and it guarantees Eq.~(\ref{eq:const eq in rz}) to be satisfied on the horizon. 
Assuming the surface gravity is nonzero (nonextremal), we find that 
$T_{,\rho}$ must be nonzero on the horizon; $T_{,\rho} \neq 0$. 
Then, from the regularity of Eq.~(\ref{eq:pol-R}) or Eq.~(\ref{eq:pol-C}) on the horizon, a condition for $C_{,\rho}$ is derived as 
\begin{eqnarray}
    C_{,\rho} = - \frac{\ell}{\rho_h z} 
    \quad (\mbox{at}~\rho=\rho_h ) \,.
\label{B.C. C at horizon}
\end{eqnarray}
On the other hand, the condition that the expansion is zero on the horizon gives~\cite{Eardley:1997hk}
\begin{eqnarray}
 (R+2C)_{,\rho} +\frac{3\ell}{\rho_h  z} = 0 
 \quad (\mbox{at}~ \rho=\rho_h ) \,, 
\end{eqnarray}
which guarantees the constraint equation (\ref{eq:const eq in zz}) on the horizon. 
In the following numerical calculation, we basically impose these
Neumann boundary conditions on the horizon for $R$ and $C$. 
As a supplementary condition to improve the accuracy, we also use 
Eq.~(\ref{R horizon}), but we do not use Eq.~(\ref{surface grav}). 
Hence, in our calculation the surface gravity is not given by hand as a parameter. 
We discuss this point in more detail at the end of this subsection.

For the extremal case $\kappa=0$, the boundary conditions are not fully determined in the present way. From $\kappa=0$, $R=\infty$ or
$T_{,\rho}=0$ are required on the horizon, and thus the boundary condition for $C$ is not determined from the regularity on the horizon. It might be interesting to explore the extremal case separately, but we do not consider it in this paper.

Israel's junction condition for the RS braneworld model gives  
$ K_{\mu\nu} = - \gamma_{\mu \nu}  / \ell $,
where $\gamma_{\mu\nu}$ and $K_{\mu\nu}$ are the induced metric and extrinsic curvature on the brane, respectively. 
From this condition, we obtain the boundary conditions on the brane as 
\begin{eqnarray}
 \frac{\partial_{\xi}T}{T} 
 = \partial_{\xi}R   
 = \partial_{\xi}C  
 = -\frac{\rho }{2\ell \sqrt{\xi} } \left(1- e^R \right) 
    \quad \left(\mbox{at}~ \xi=(\pi/2)^2\right). 
\end{eqnarray}
The constraint equation (\ref{eq:const eq in rz}) is manifestly satisfied under these conditions.

We must also specify asymptotic boundary conditions.
For the asymptotic infinity, the boundary conditions to obtain asymptotically AdS spacetime are $T\to 1$ and $R, C \to 0$. 
The metric functions must smoothly approach these asymptotic values. In actual numerical calculations, these asymptotic boundary conditions are imposed at a finite, but sufficiently far region, and we must check that the solutions are insensitive to the position.

With these boundary conditions, we can solve the elliptic equations as a boundary value problem. Moreover, these boundary conditions guarantee the constraint equation $\Theta_1$ to be satisfied on all boundaries and $\Theta_2$ to be satisfied at least on the horizon. 
Hence as we explained, the two constraint equations are automatically satisfied as long as the elliptic equations are exactly solved.

It should be noted that all elliptic equations, constraint equations, and boundary conditions can be rewritten in terms of nondimensional coordinates $\{\hat{\rho},\chi\}$,  where $\hat\rho = \rho/\ell$, without any dimensionful parameters.
Then this system of equations, and hence a black hole solution, is characterized only by a single parameter 
\begin{eqnarray}
{L} = \ell/\rho_h. 
\label{eq: def L}
\end{eqnarray}
In our numerical calculation, we take $\rho_h \equiv 1$ and change
$\ell$ to specify this parameter, since we want to fix a coordinate region where numerical calculations are performed. 
The variation of $\ell$ keeping $L$ fixed corresponds to a rescale of the length scale as we notice from the metric (\ref{eq:assume}), which is rewritten as 
\begin{eqnarray}
ds^2 = \ell^2 ds^2_{L} (\hat{t},\hat\rho,\chi)  ,  
\end{eqnarray}
where $ds^2_{L}$ is the nondimensional part of the line element written by
$\{ \hat{t} = t/\ell, \hat\rho, \chi \}$, and its metric function is given by a black hole solution $T, R$, and $C$ with the parameter $L$. 
After specifying a black hole solution by this method, one may want to transform the solutions to those in the coordinates where $L$ is specified by taking $\ell\equiv 1$ and changing $\rho_h$. For this purpose, it is enough to multiply $\ell^{-1}$ to a length scale in the coordinates where $\rho_h \equiv 1$ is taken.

A question may arise. Although we say that we use ${L} = \ell/\rho_h$ to specify a solution, $\rho_h$ does not have a clear physical meaning. One may think that it is more appropriate to specify the value of the surface gravity $\kappa$ instead of $\rho_h$. 
Here we should recall the argument given at the end of the preceding subsection that $\rho/\rho_{h}$ and hence $\hat\rho/\hat\rho_{h}$ cannot be changed arbitrarily by using the residual gauge degrees of freedom. 
If we change the value of $\hat \rho_h$ by using this residual gauge degree of freedom, the value of $\hat\rho$ at infinity also scales correspondingly. Then, we will see that the asymptotic boundary conditions $T\to 1$ and
$R, C \to 0$ are not satisfied after the gauge transformation. 
Hence, for a given solution we do not have a degree of freedom to change the value of $\hat\rho_h$ arbitrarily. Thus, if we specify a solution by fixing both $\kappa$ and $\ell$, we are not allowed to set $\rho_h=1$ any further. 
On the contrary, as we wish to set $\rho_h=1$ in the actual numerical computation, we should not use the condition 
(\ref{surface grav}) that specifies the value of $\kappa$ directly.

\section{Numerical examples of small black holes}
\label{sec:3}

\subsection{Numerical examples}

\begin{figure}
\includegraphics[width=9cm,clip]{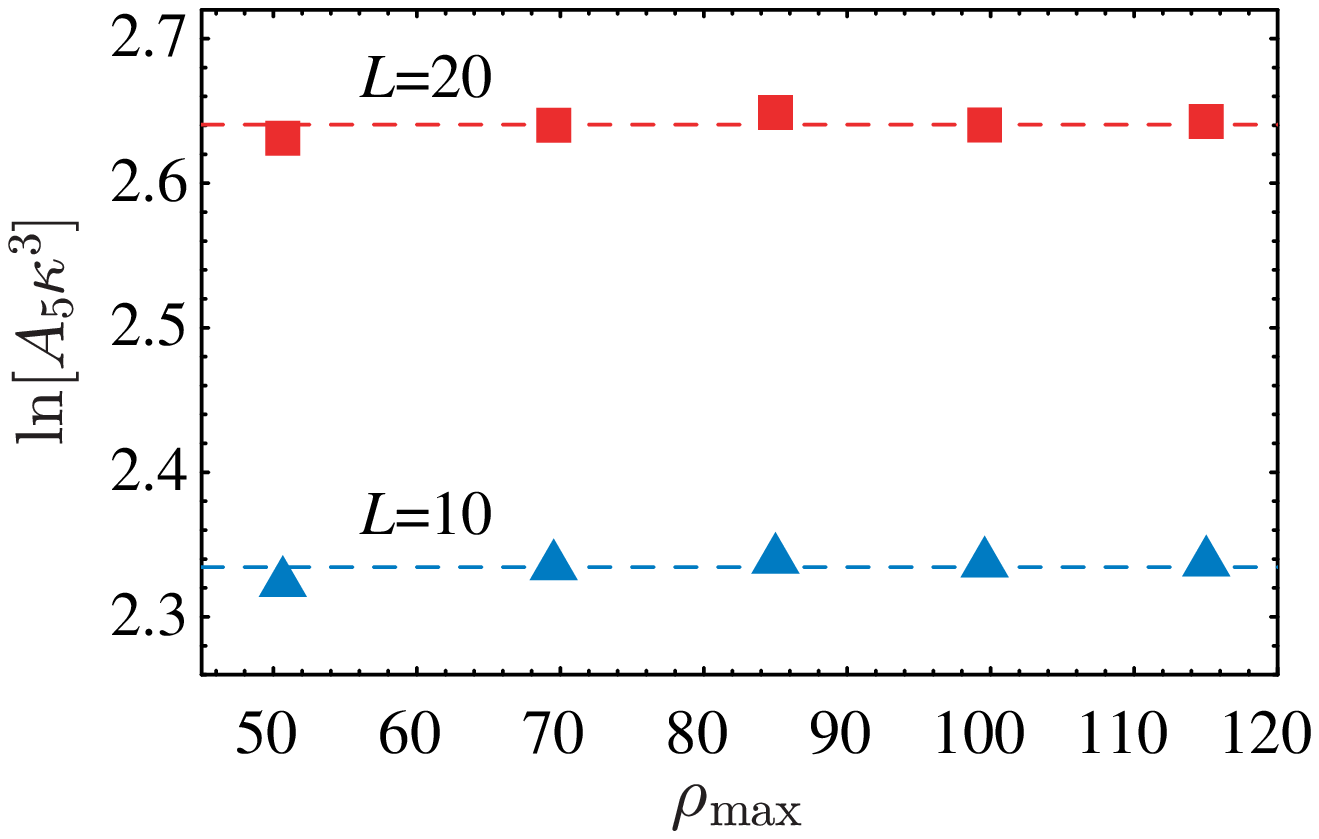}
\caption{
\label{fig:2ndoc}
A plot showing the sensitivity of the numerical solutions to the position of asymptotic boundary. The calculations are performed for $L=10$ and $L=20$. 
The position of asymptotic boundary $\rho_{\mathrm{max}}$ is changed from $\rho_{\mathrm{max}}=50$ to $\rho_{\mathrm{max}}=115$ keeping the ratio of the resolution fixed at $\delta\rho_{\mathrm{max}}/\delta\rho_1 \approx 2.71$ with $\delta\rho_1 \approx 0.05$.  
We see that the variation of the thermodynamic quantity is no more than  $1\%$, and is stable for this parameter range of $\rho_{{\mathrm{max}}}$. 
}
\end{figure}

The formulation outlined in the previous section indeed works well to solve the Einstein equations as a boundary value problem, using an iterative convergence scheme of the relaxation method. 
A simple check of an algorithm is to calculate a very large limit of $L$ for fixed $\rho_h$, and compare the result with the 5D Schwarzschild black hole in isotropic coordinates which is a solution in the small horizon limit $L \to \infty$ $(\ell \to \infty)$ taking $z/\ell \to 1$
\begin{eqnarray}
T(\rho) &=& \frac{\rho^2 - \rho_h^2}{ \rho^2+\rho_h^2}. 
\cr
R(\rho)&=&C(\rho) = \log \left(1+ \frac{\rho_h^2}{\rho^2} \right). 
\label{5D Schwarzschild}
\end{eqnarray}
In this limit, the tension of the brane vanishes and the background spacetime is flat. 
We have used this metric as an initial guess configuration with which we start the relaxation. 
Namely, we first calculated a solution for relatively large $L$, and then proceeded to smaller $L$ by using the result relaxed for the previous value of $L$ as the new initial configuration.
Convergences of numerical calculation become worse as $L$ becomes smaller, and errors of constraint equations also grow if we keep the same resolution. 
It becomes in general more difficult to find solutions as the nonlinear effects of differential equations grow, and we could not keep the convergence of the calculation for $L \lesssim 1$. 
Thus in this paper we show only small black holes ($L > 1$), although it is important to improve our scheme by identifying and removing all numerical instabilities.
The number of grid points ($\xi_i, \rho_j$) used in our calculation is $100 \times 1000$. The grid point $\rho_j$ is taken to be a geometric progression $\rho_j=\rho_h + \epsilon (1- \gamma^j)/(1-\gamma)$.
The parameters $\epsilon$ and $\gamma$ are determined by requiring that the ratio of the last grid resolution to the first one is $\delta\rho_{\mathrm{max}}/\delta\rho_1\approx 2.71$ and the asymptotic boundary is set at $\rho_\mathrm{max}=85$.
It is crucial to test the sensitivity of the solutions to the physical size of the lattice. This will be discussed later (see Fig. \ref{fig:2ndoc}).

\begin{figure}
\includegraphics[width=7cm,clip]{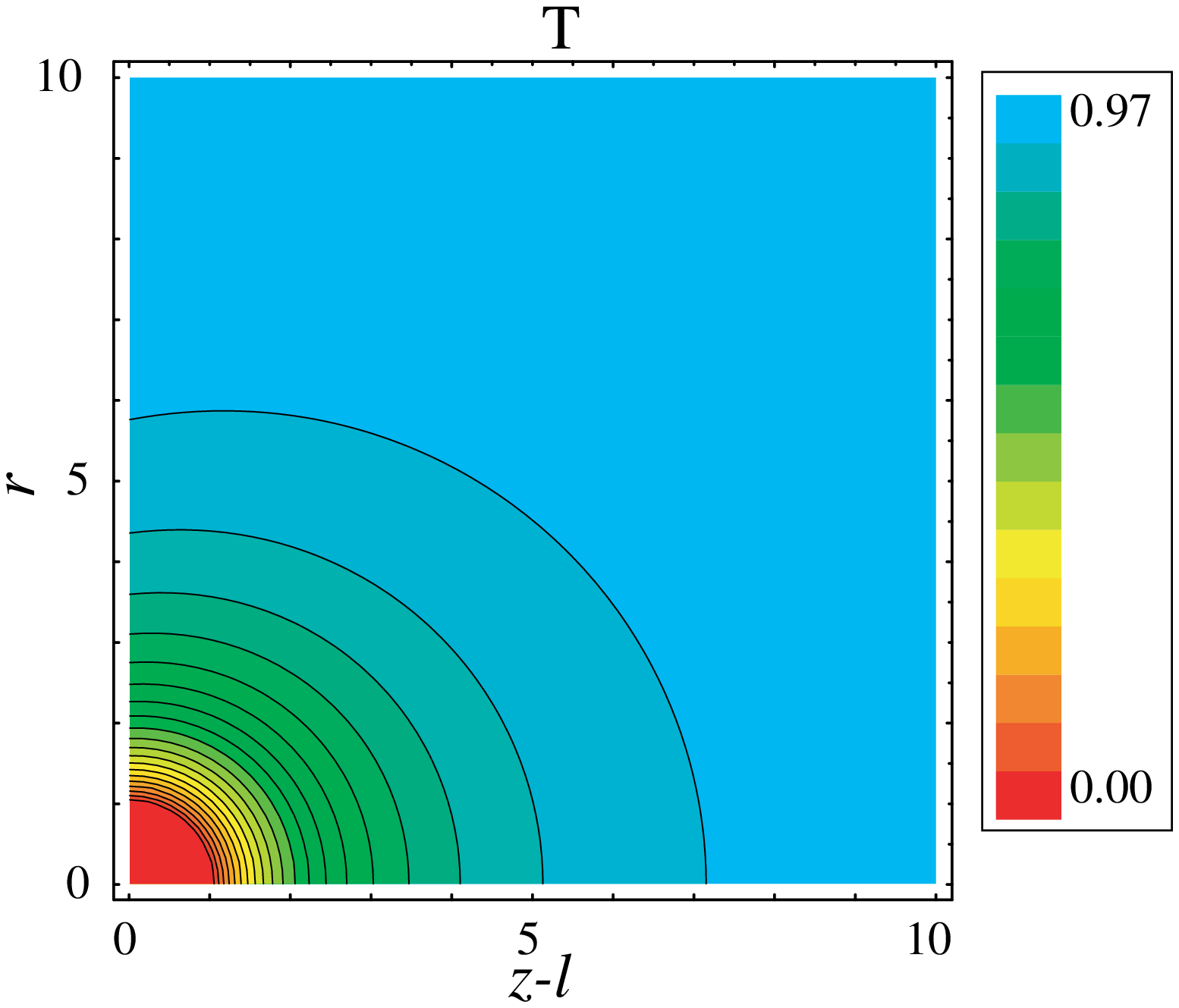}
\includegraphics[width=7cm,clip]{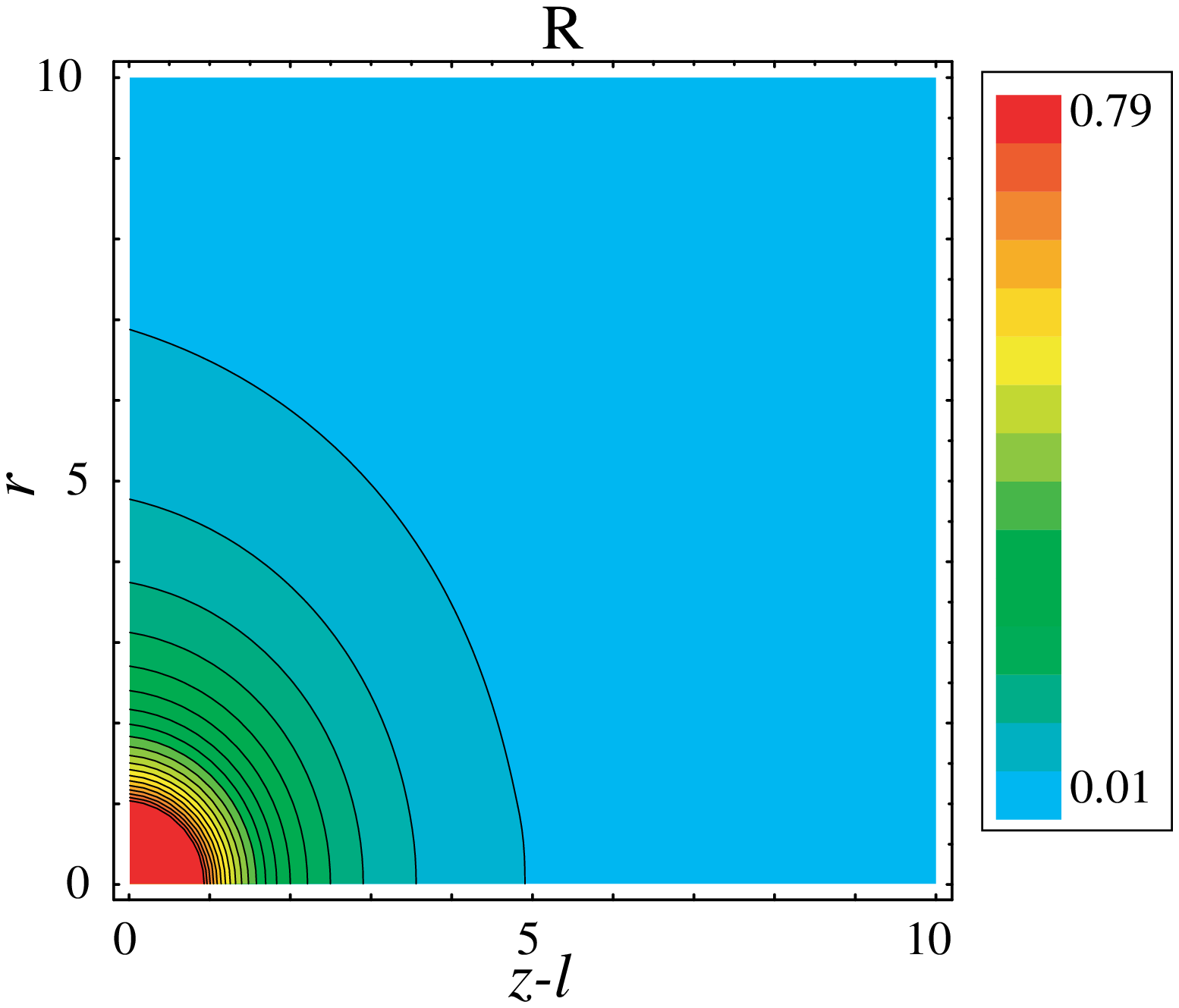} 
\includegraphics[width=7cm,clip]{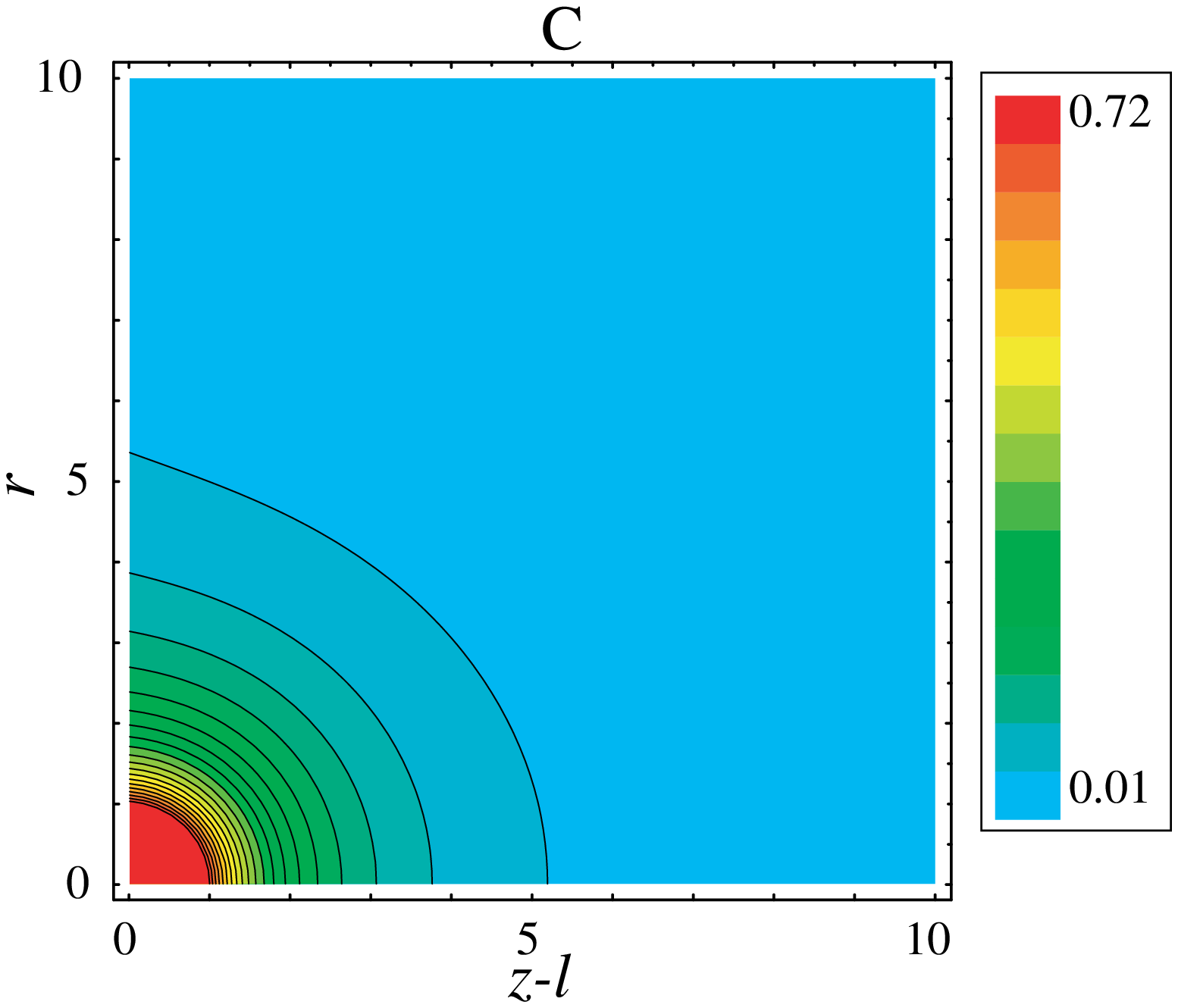} 
\caption{
\label{fig:L15TRCrz}
An illustration of the metric functions $T$, $R$, and $C$ for $L=15$ in the coordinates $\{r,z \}$. 
This is a typical example of our numerical solutions for small black holes ($L \gtrsim 5$). 
In these figures the center of the black hole is placed at the bottom left corner ($r=0$ and $z-\ell=0$), and the brane is at $z-\ell=0$. 
The numerical calculation has been performed in the polar coordinates
$\{ \rho, \xi \}$, and hence the inner region of the black hole $\rho<\rho_h$ is outside the area of our computation.
 }
\end{figure}

To orient the reader, we show the result of $T, R$, and $C$ for
$L=15$ in Fig. \ref{fig:L15TRCrz} as a typical example.
The center of the black hole is at $r=0$ and $z-\ell=0$.
The interior of the horizon $\rho< \rho_h$ is outside the region of our numerical calculation. The contours of $T$ are almost spherical for this value of
$L$ as in the case of the limit $L\to\infty$ in spite of the nontrivial boundary conditions on the brane, while the contours of $R$ and $C$ manifestly deviate from spherical shape.

\begin{figure}
\includegraphics[width=7cm,clip]{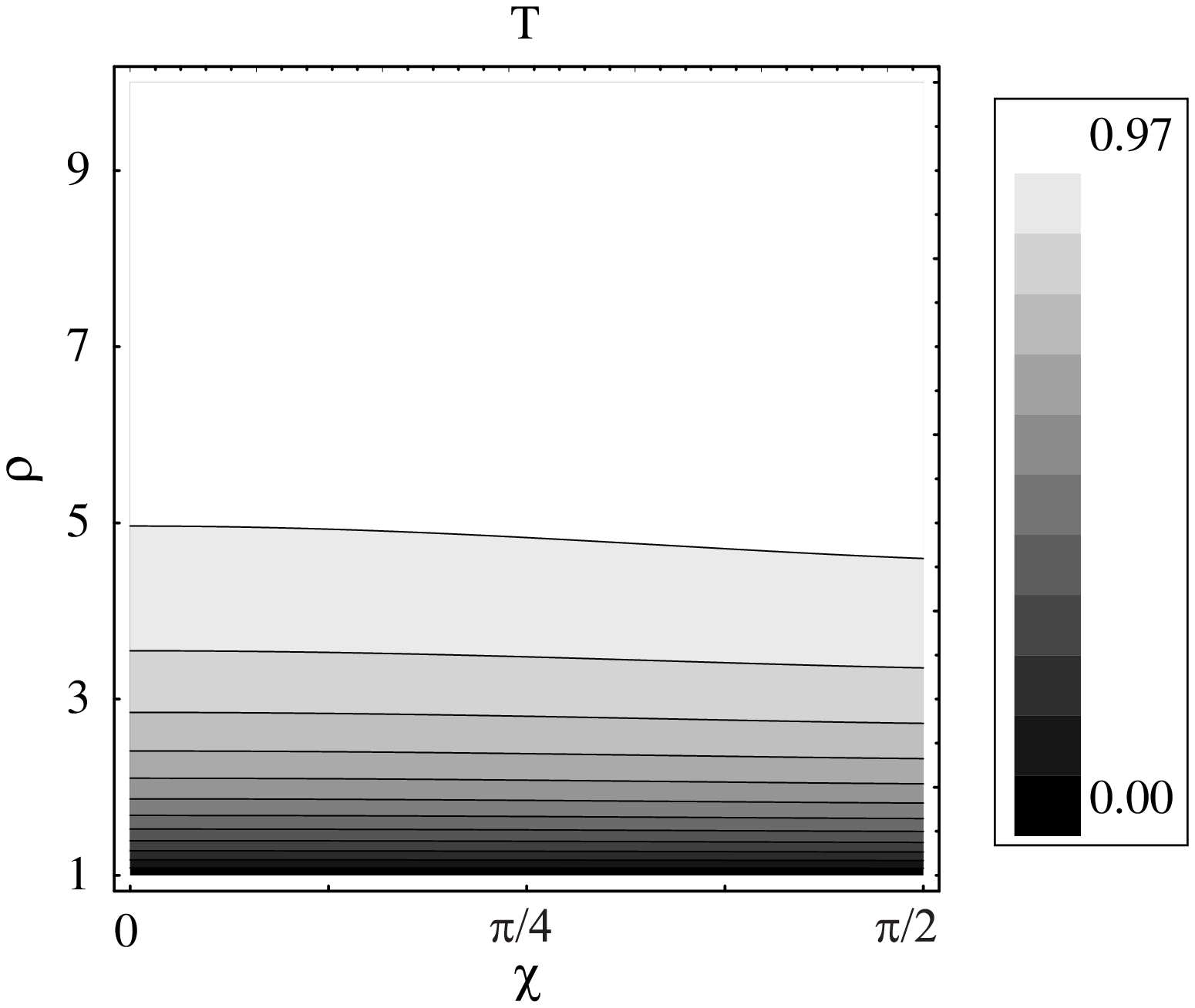} 
\includegraphics[width=7cm,clip]{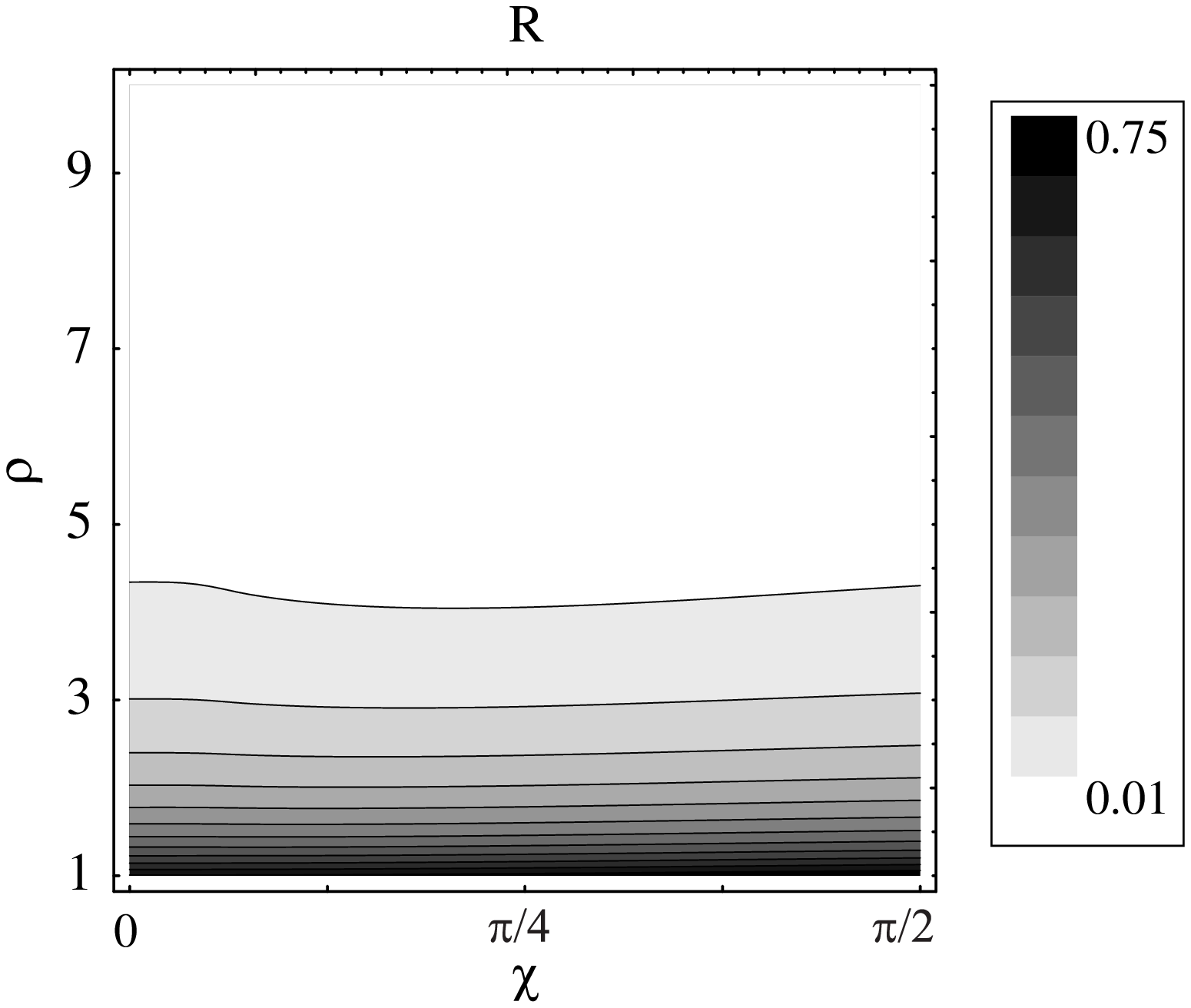} 
\includegraphics[width=7cm,clip]{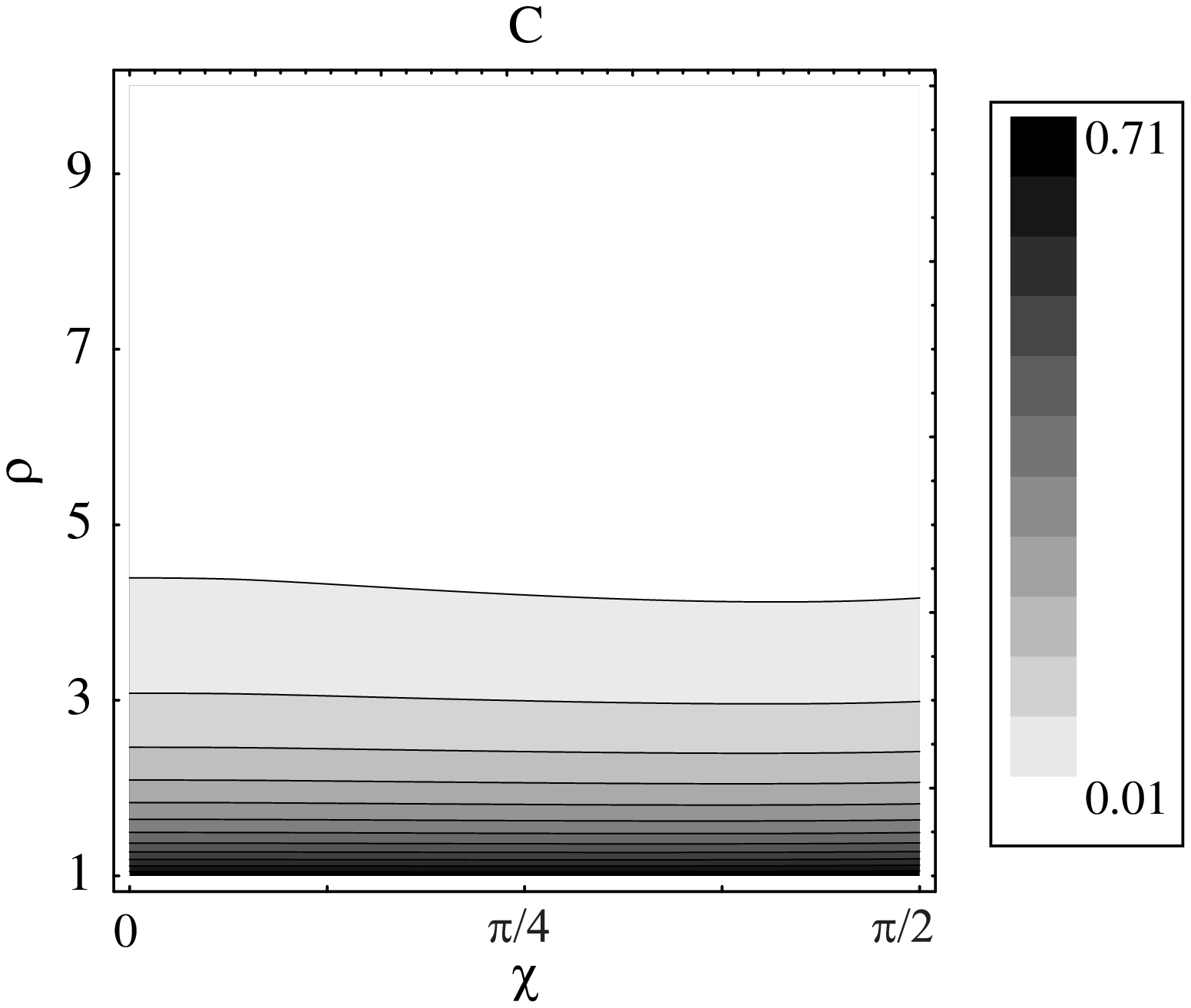} 
\caption{
\label{fig:L30TRC}
An illustration of the solution for $L=30$ in polar coordinates $\{\rho,\chi\}$. 
} 
\end{figure}

\begin{figure}
\includegraphics[width=7cm,clip]{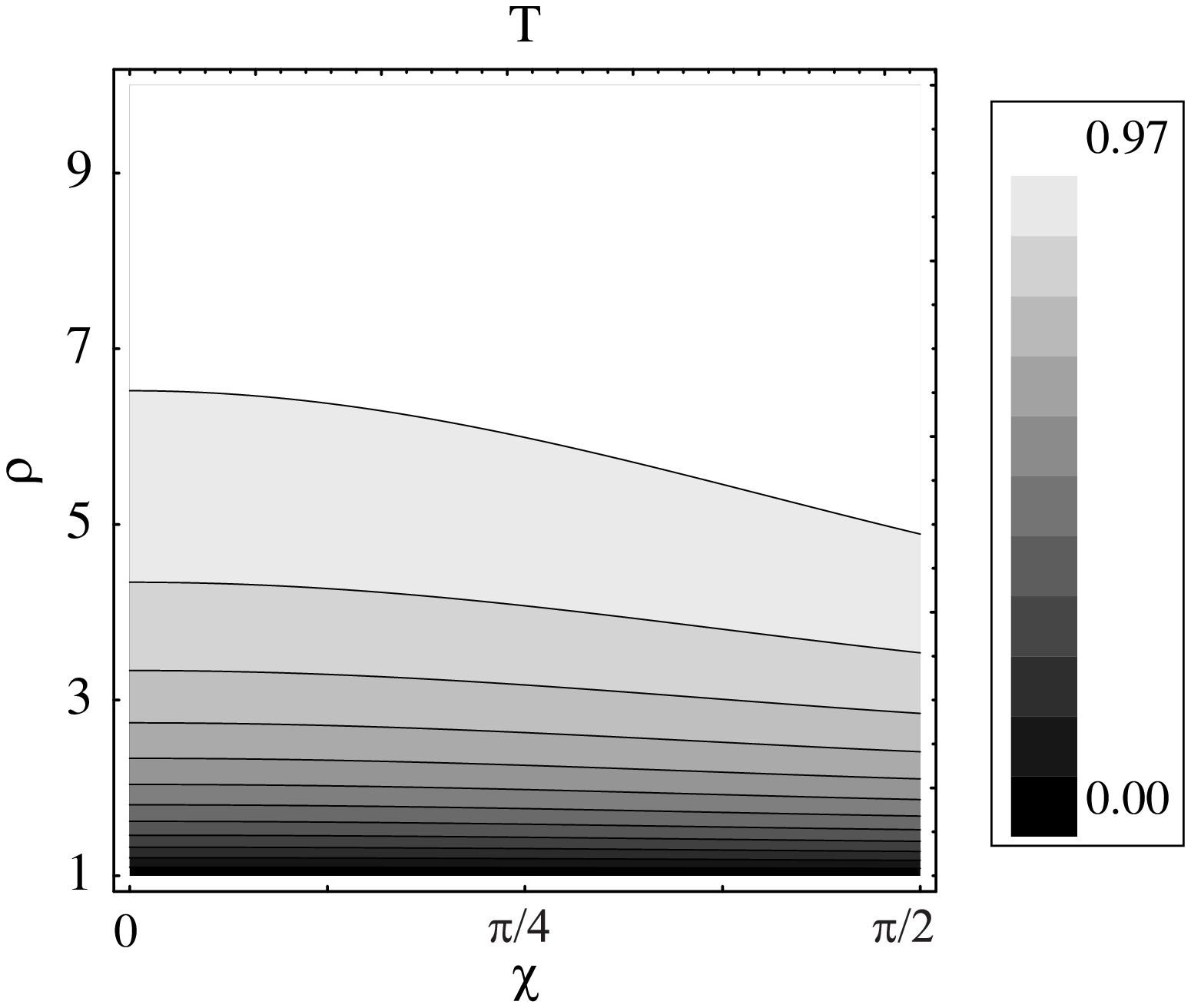} 
\includegraphics[width=7cm,clip]{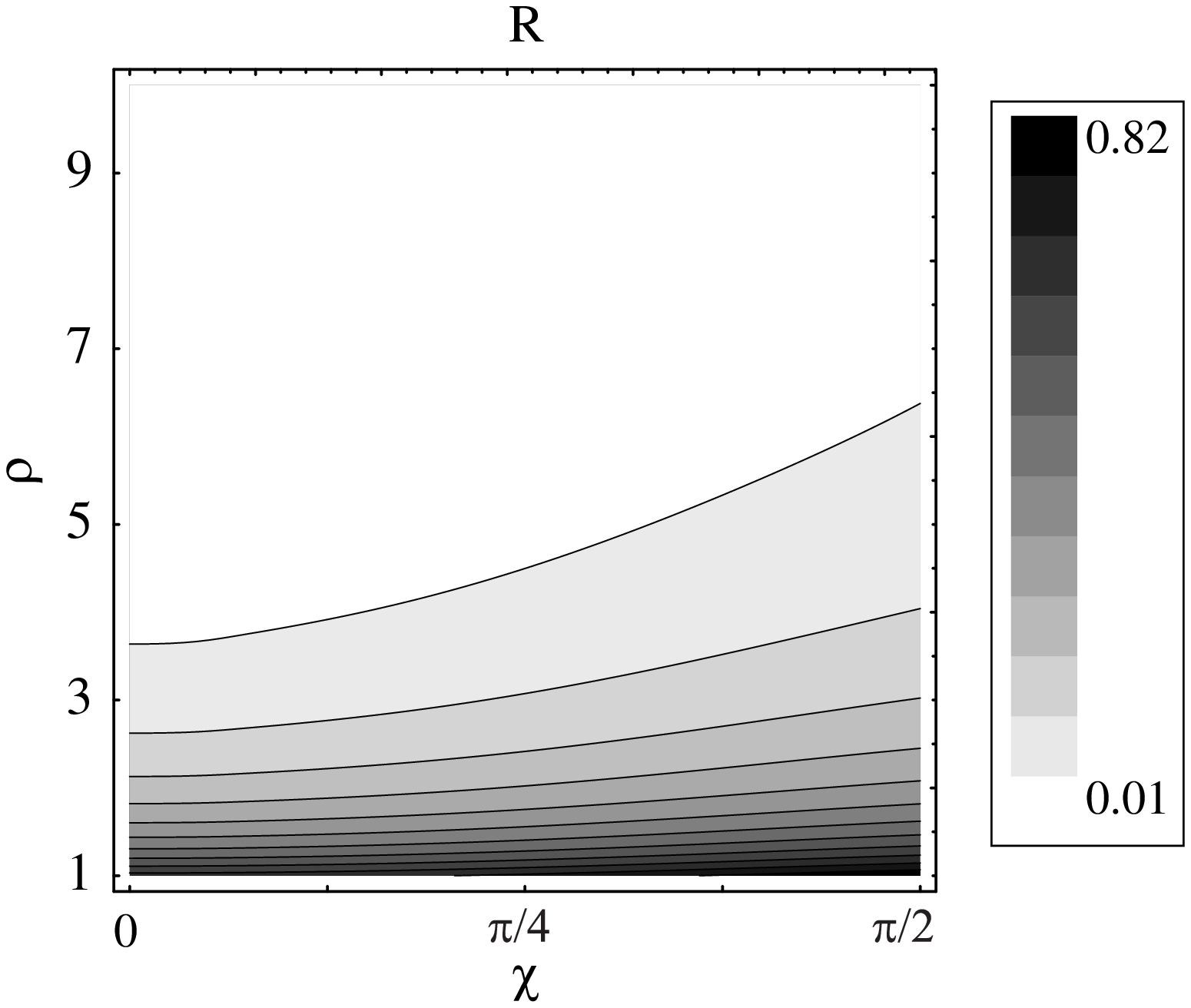} 
\includegraphics[width=7cm,clip]{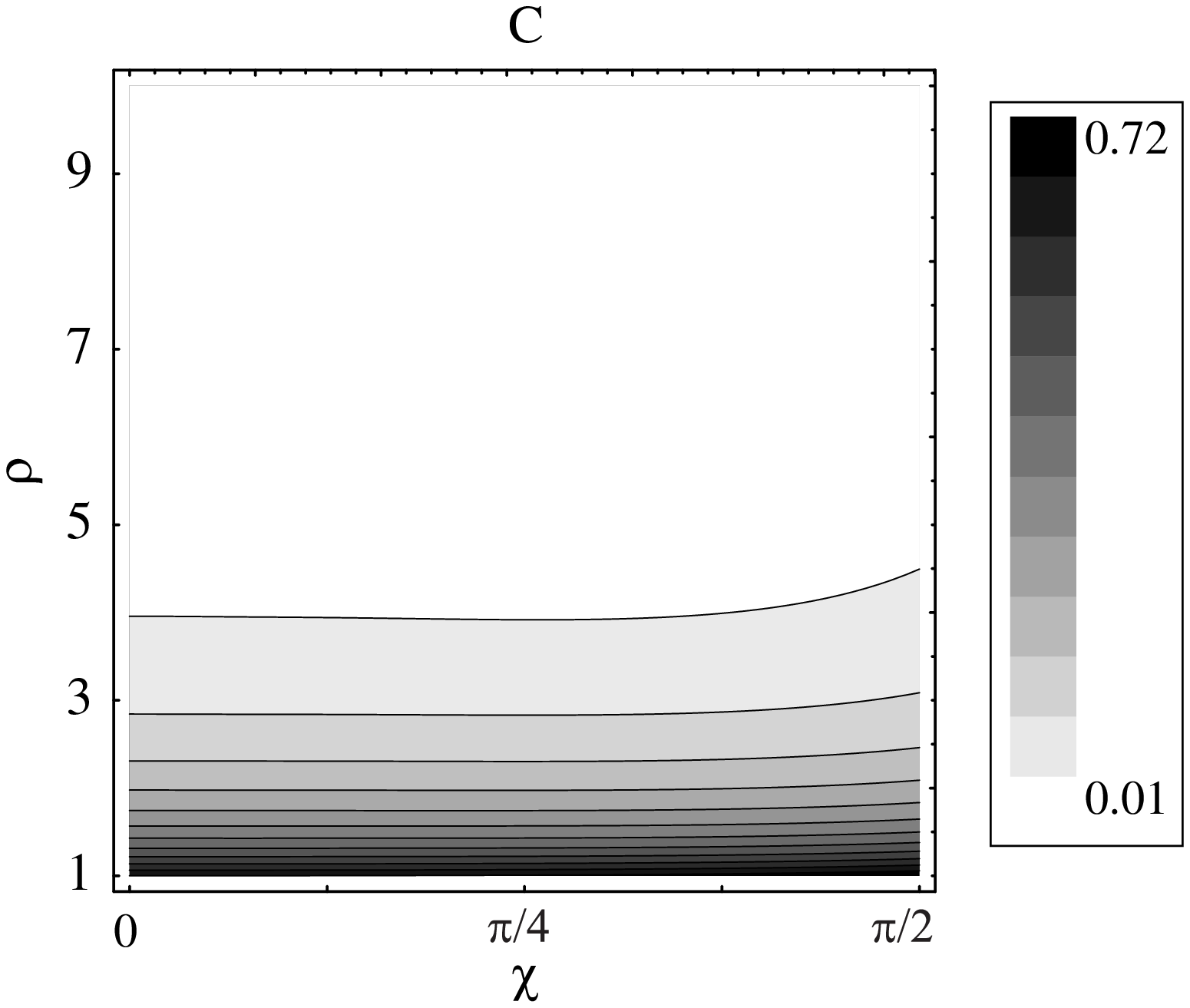} 
\caption{
\label{fig:L10TRC}
An illustration of the metric functions $T$, $R$, and $C$ for $L=10$ in the coordinates $\{\rho, \chi\}$. 
}
\end{figure}

In Fig. \ref{fig:L30TRC}, the solution for $L=30$ is displayed in the polar coordinates that are more appropriate to see the angular dependence of the metric functions.
Figure \ref{fig:L10TRC} displays the results for $L=10$ in the polar coordinates. The deviations from spherical shape for $L=10$ are enhanced more than those for $L=30$. 
We have performed numerical calculations for the parameter region $ L=3 \sim 500$. The results for some $L$ are listed in Table \ref{table:data}.

To consider the shape of the small localized black hole in the bulk, we calculated geometrical quantities for the numerically obtained black hole solutions.  The area of the black hole horizon is given by 
\begin{eqnarray}
 { A}_5 
 =  2 \rho_h^3\int d\Omega \int^{({\pi}/{2})^2}_0 
    d\xi  \frac{ \sin^2 \sqrt{\xi}}{2\sqrt{\xi}}
    \left( \frac{\ell}{z}\right)^3 e^{R+2C}  .
\label{BH area}
\end{eqnarray}
Here the factor 2 is due to $Z_2$ symmetry.  
The proper area of the intersection between the horizon and the brane is given by 
\begin{eqnarray}
 A_4 = 4\pi \rho_h^2  e^{2C} \quad (\chi=\pi/2).
 \label{4D area}
\end{eqnarray}

Figure \ref{fig:A4A5} shows the ratio of a mean radius in four-dimension  $\sqrt{A_4}$ (on the brane) to that in five-dimension $A_5^{1/3}$. 
 The figure indicates that the black hole tends to flatten as its horizon radius increases.  This geometrical behavior is expected from the AdS geometry of whole spacetime. The so-called warp factor $\ell z^{-1}$ of the AdS geometry (\ref{eq:assume}) will in general cause the localized black hole to be flattened, and this effect becomes significant when the horizon radius is large. 
In fact, the EHM solution exhibits a similar behavior to Fig.~\ref{fig:A4A5}.

\subsection{Constraint equations and checks of calculation} 
\label{sec:constraint and checks}

\begin{figure} 
\includegraphics[width=7cm,clip]{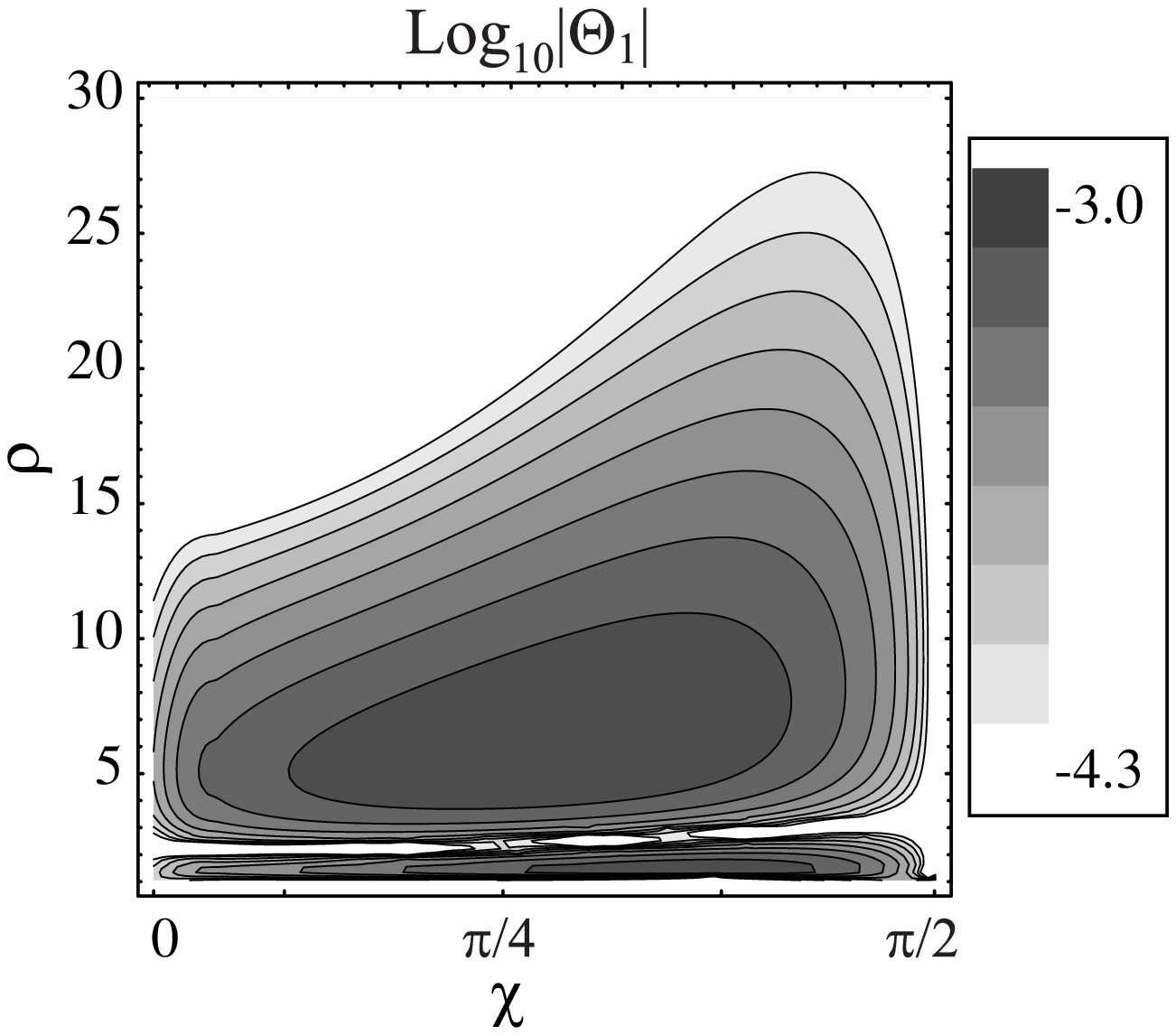}
\includegraphics[width=7cm,clip]{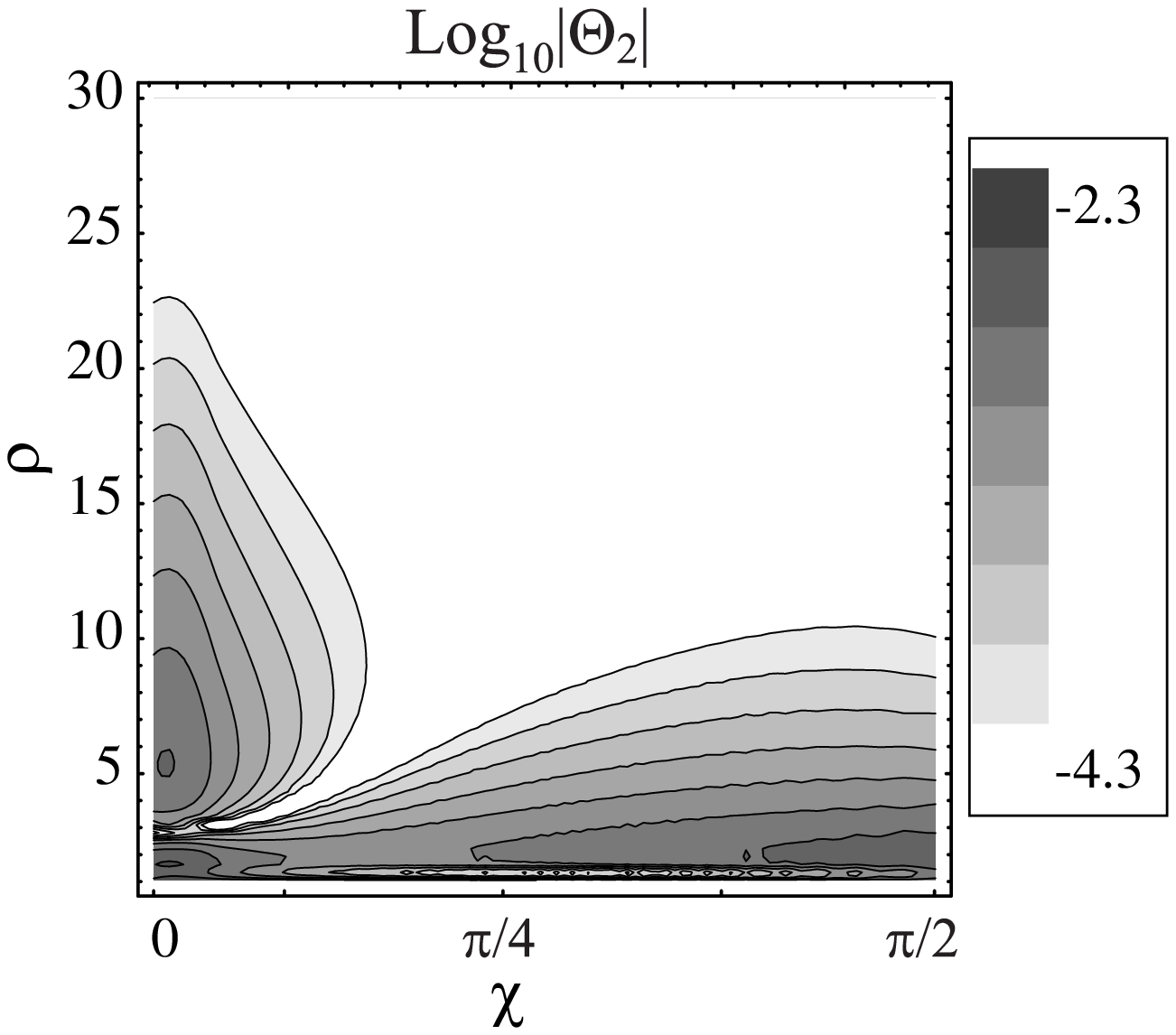} 
\includegraphics[width=7cm,clip]{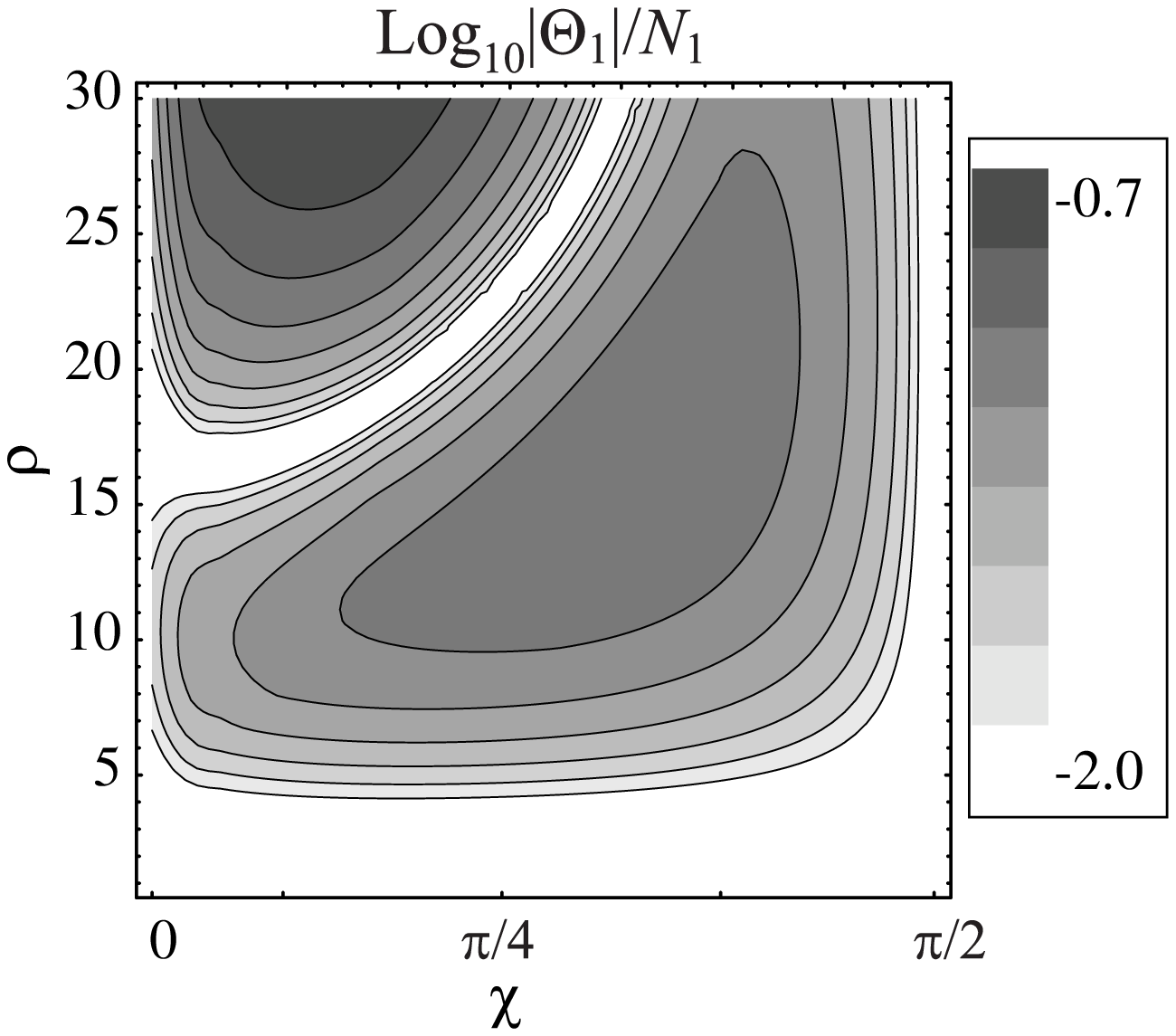} 
\includegraphics[width=7cm,clip]{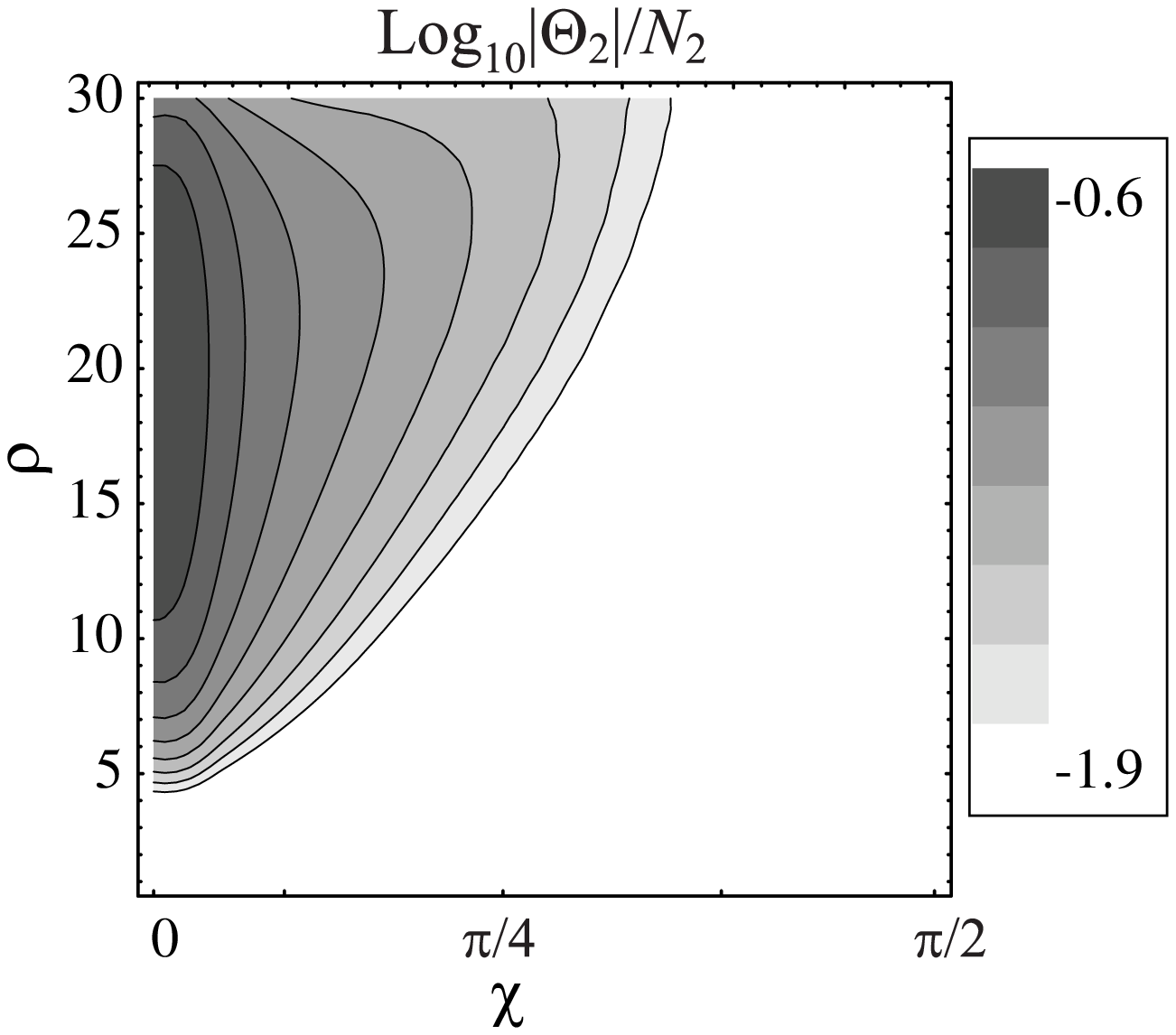} 
\caption{
\label{fig:Constraints}
An illustration of the constraint equations for $L=15$. 
The figures that plot the absolute values of two constraint equations, $|\Theta_1|$ and $|\Theta_2|$, show the absolute errors of the constraint equations. The other two figures plotting $|\Theta_1(\rho,\chi)|/N_1 (\rho,\chi)$ and $|\Theta_2(\rho,\chi)|/N_2 (\rho,\chi)$ show the relative accuracy of the constraint equations since $N_1$ and $N_2$ are norms of respective constraint equations (see Table \ref{table:data}). 
As expected, the absolute errors of the constraint equations are observed mainly near the axis of polar coordinates and around the horizon.
However, the relative accuracy is not significantly low around the horizon, but it is worse near the axis and in the asymptotic region.
}
\end{figure}

In this section we discuss the validity of our numerical calculations.
We have calculated the constraint equations for each numerical solution.
Averaged absolute errors of the elliptic equations and the constraint equations are listed in Table \ref{table:data}. In the table, we have introduced a norm $N(\xi,\rho)$ of the constraint equations since we need appropriate references to compare with the violations of constraint equations. 
The norm is defined by 
\begin{eqnarray}
N (\xi, \rho)=\sum_k |n^{(k)} (\xi, \rho)| \,,
\label{norm}
\end{eqnarray}
where $n^{(k)}$ represents the terms in $\Theta_1$ or $\Theta_2$ at $(\xi,\rho)$.
The comparison between this norm and the violation of the constraint equation gives a degree of cancellation between different terms.
We see in Table \ref{table:data} that averaged constraint violations against averaged norms are small, and thus the constraint equations appear to be well satisfied.
Moreover by using this norm, we can obtain relative accuracies of the constraint equations, $|\Theta_1|/N_1$ and $|\Theta_2|/N_2$.
Figure \ref{fig:Constraints} shows that the absolute errors of the constraint equations, $|\Theta_1|$ and $|\Theta_2|$, are observed mainly around the origin of polar coordinates and near the horizon where the elliptic equations have terms whose coefficients behave like $1/\xi \to \infty$ or $1/T \to \infty$.
On the other hand, the figure represents that the relative accuracy around the horizon is not so bad, although the relative accuracy becomes worse near the axis.  This violation of the constraint equations is a general feature of the present calculations and it becomes larger as $L$ becomes smaller.
The lack of accuracy near the axis is a common problem in axisymmetric problems.

 We have also checked second-order convergences of the thermodynamic quantities for $L=20$ keeping $\rho_{\mathrm{max}}=85$.
As a function of $\delta \xi$, we confirmed the quadratic fits for $\kappa$ and $A_5$. 
The similar quadratic fit is possible for the variation of $\rho$ resolution, in which we need to change the lattice resolution keeping the ratio $\delta\rho_{\mathrm{max}}/\delta\rho_1$ fixed ($\approx 2.71$). 
The numbers of grid points $(\xi_i,\rho_j)$ used in these checks are $(80-200)\times 1000$ for $\delta\xi$ variation, and $100 \times (750-2000)$ for $\delta\rho$ variation, which correspond to $\delta\xi = 0.01 - 0.03$ and $\delta\rho = 0.02 - 0.07$, respectively.
Note that the absolute errors decrease as the lattice resolutions increase. However, clear second-order convergence has not been observed because we have not used equal grid spacing for $\rho_j$. If we use equal grid spacing, second-order convergence is actually observed.
We have also checked the insensitivity of our numerical solutions on the finite position of the asymptotic boundary. To confirm it, we vary the position $\rho_\mathrm{max}$ for $L=10$ and $L=20$, and check the robustness of thermodynamic quantities. 
For each variation of $\rho_\mathrm{max}$, we have changed the grid resolution for $\rho$ keeping the ratio $\delta\rho_{\mathrm{max}}/\delta\rho_1 \approx 2.71$ with $\delta \rho_1 \approx 0.05$.
The thermodynamic quantities are stable for each value of $\rho_{\mathrm{max}}$. Figure \ref{fig:2ndoc} shows a result for a nondimensional combination. One sees that the combination of thermodynamic quantities is stable and the variation is no more than $1\%$ for the parameters.

\section{ Summary } 
\label{sec:Discussion} 

\begin{figure}
\includegraphics[width=9cm,clip]{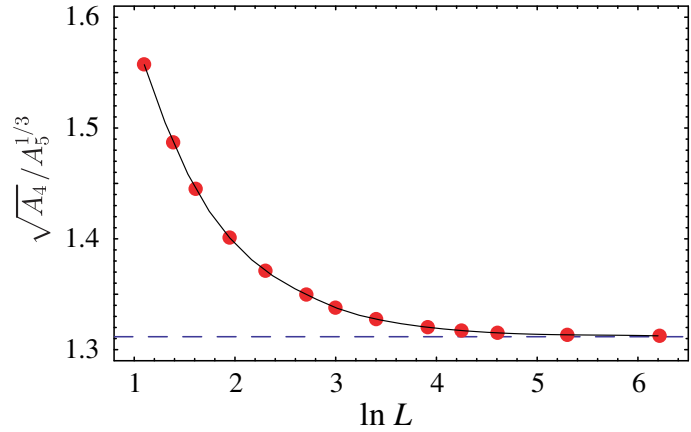}
\caption{
\label{fig:A4A5}
An illustration of $\sqrt{A_4}/A_5^{1/3}$. This plot shows a degree of deformation of a black hole. 
One sees that as $L$ decreases the ratio of a mean radius in four-dimension  ${A_4}^{1/2}$ (on the brane) to that in five-dimension $A_5^{1/3}$ increases. 
It indicates that the black hole tends to flatten as its horizon radius increases.  
The dashed line shows the same quantity for the 5D Schwarzschild black hole. Note that $L$ is expressed in natural logarithm.
}
\end{figure}

Black hole solutions that represent a black hole localized on the brane in the Randall-Sundrum infinite braneworld model have not been found with appropriate boundary conditions in the literature. We have explored this problem in this paper.
We have performed the numerical calculations and found nontrivial localized black hole solutions whose horizon radii $\rho_h$ are small compared to the bulk curvature scale $\ell$. 
More precisely, 
\textit{the small localized black hole solutions are approximately  constructed by numerical calculations under appropriate boundary conditions.}  
From the numerically obtained solutions, we could observe the transition of the shape of the small localized black hole in the bulk (Fig. \ref{fig:A4A5}).  The black hole tends to flatten as its horizon radius increases. This is expected due to the nature of the AdS geometry, and is also observed in the EHM solution of a localized black hole in the 4D braneworld~\cite{Emparan:2000wa}. 
Although the EHM solution is the solution in lower dimensions, it is interesting to compare our solution with it in more details.

Although the method we have developed works well for the small localized black holes, we could not succeed in finding black hole solutions with large horizon radius. 
The lack of convergence in the relaxation scheme is due to the appearance of unstable modes. For example, an unstable mode that appears near the axis is a common problem in axisymmetric code (see, e.g., \cite{Choptuik:2003}). In particular, the first terms in Eqs. (\ref{eq:pol-R}) and (\ref{eq:pol-C}) that have the factor $1/r^2$ give severe contributions near the axis, and they invoke numerical instability.
The same problem appeared in the calculation of a relativistic star in the braneworld model \cite{Wiseman:2001xt}.
Thus improvement of the numerical scheme and formulation to avoid this instability is an important issue.
It may be worth pointing out that since we do not have any uniqueness theorem in the context of the braneworld~\cite{Gibbons:2002av,Gibbons:2002ju}, there is a possibility of finding a sequence of solutions other than that found in this paper if we start with a completely different initial guess.

It is worthwhile to point out that the boundary conditions that we have imposed do not {\it  a priori} guarantee the absence of naked singularities in the bulk or on the horizon.
If any solution obtained in our formulation cannot avoid a naked singularity in the bulk or on the horizon, it means that there is no physically acceptable static black hole in the RS infinite braneworld. Such a possibility was pointed out for large localized black holes~\cite{Tanaka:2002rb,Emparan:2002px}.

As a direction of future work, it is crucially important to find large black hole solutions if they exist~\cite{Tanaka:2002rb,Emparan:2002px}, and it needs further developments of technique and investigation.  It is also interesting to apply this method to find the black holes localized on the TeV brane~\cite{Randall:1999ee}. 
The stabilities of a higher-dimensional black hole including our solution are also important.   We hope the method used in this paper will become the basis for further studies on black holes in higher dimensions.

\begin{acknowledgments}
We would like to thank Shu-ichiro Inutsuka, Toby Wiseman, Tetsuya Shiromizu, Roberto Emparan, Nemanja Kaloper, and Shinji Mukohyama for their helpful comments and suggestions. 
To complete this work, the discussion during and after the YITP workshops YITP-W-01-15 and YITP-W-02-19 were useful. 
Numerical computations in this work were carried out at the Yukawa
 Institute Computer Facility.
H.K. is supported by the JSPS.
This work is partly supported by the Monbukagakusho Grant-in-Aid No. 1270154, 14047212 and 14204024.
\end{acknowledgments} 
 
\appendix

  

\begin{table}
\begin{tabular}{cccccccccc}
\hline   \hline 
$L$  &$\left\langle|TRC|\right\rangle$ & $\left\langle |\Theta_1|\right\rangle$ & $\left\langle |\Theta_2|\right\rangle$  
     & ($\left\langle N_1 \right\rangle$, $\left\langle N_2 \right\rangle$)
     & $\kappa$  &  ${A}_5$  &  ${A}_4$& 
\\
\hline
 5 &5.9\e{-4}&6.8\e{-4}&6.8\e{-4}&(2.2\e{-2}, 3.7\e{-2})&0.37&129&53&\\ 
10 &1.3\e{-4}&2.2\e{-4}&1.6\e{-4}&(1.2\e{-2}, 2.7\e{-2})&0.41&150&53&\\ 
15 &9.7\e{-5}&2.5\e{-4}&1.4\e{-4}&(9.2\e{-3}, 2.5\e{-2})&0.43&155&53&\\ 
30 &7.4\e{-5}&1.7\e{-4}&9.3\e{-5}&(5.8\e{-3}, 2.4\e{-2})&0.46&161&52&\\ 
50 &6.3\e{-5}&1.2\e{-4}&6.2\e{-5}&(4.8\e{-3}, 2.4\e{-2})&0.47&160&51&\\ 
100&5.0\e{-5}&6.9\e{-5}&5.6\e{-5}&(4.1\e{-3}, 2.3\e{-2})&0.48&158&51&\\ 
500&9.8\e{-6}&1.4\e{-5}&1.7\e{-5}&(3.6\e{-3}, 2.3\e{-2})&0.50&158&50&\\ 
\hline 
$L= \infty$&  &      & &    &0.50&158&50\\
\hline \hline   
\end{tabular}
\caption[short]{
Table summarizing the data of numerical calculations for several values of  $L=\ell/\rho_h$. In the calculations we have varied $\ell$ keeping the horizon radius $\rho_h=1$. Thus it should be noted that the dimensionful parameters, i.e., surface gravity $\kappa$, 5D area $A_5$, and 4D area $A_4$, must be rescaled for comparison, as we have explained below Eq. (\ref{eq: def L}). 
In the table, $L=\infty$ represents the 5D Schwarzschild black hole, which is a solution in this limit. 
$\left\langle|\Theta_1|\right\rangle$,
$\left\langle|\Theta_2|\right\rangle$, and
$\left\langle|TRC|\right\rangle$ are averaged violations  of the constraint equations and the elliptic equations. $\left\langle|TRC|\right\rangle$ is the mean absolute value for the equations of $T$, $R$, and $C$ with equal weight for each grid point. In taking these averages, the first $5\%$ of lattice points in $\xi$ and the region near the horizon
$\rho<2\rho_h$ are excluded since they receive unphysical enhancements of errors due to singular terms.  
We introduce $\left\langle N_1 \right\rangle$ and $\left\langle N_2 \right\rangle$ as the averaged value of $N$ [Eq. (\ref{norm})] for $\Theta_1$ and $\Theta_2$, respectively, in the restricted grid region mentioned above.  
The number of grid points $(\xi_i, \rho_j)$ used in the calculations presented in this paper is $100 \times 1000$. The coordinate position of grid point $\rho_j$ is taken to be a geometric progression $\rho_j =\rho_h + \epsilon (1-\gamma^j)/(1-\gamma)$. 
The grid spaces used in the calculation are $\delta\xi=0.025$, $\delta\rho_1=0.049$,
$\delta\rho_{\mathrm{max}}/\delta\rho_1 =2.714$. 
The position of the asymptotic boundary is set at
$\rho_{\mathrm{max}}=85$. 
\label{table:data} 
}
\end{table}

\begin{thebibliography}{99}

\bibitem{Arkani-Hamed:1998rs}
N.~Arkani-Hamed, S.~Dimopoulos and G.~R.~Dvali,
 ``The hierarchy problem and new dimensions at a millimeter,''
Phys.\ Lett.\ B {\bf 429}, 263 (1998);
 I.~Antoniadis, N.~Arkani-Hamed, S.~Dimopoulos and G.~R.~Dvali,
 ``New dimensions at a millimeter to a Fermi and superstrings at a TeV,''
Phys.\ Lett.\ B {\bf 436}, 257 (1998)


\bibitem{Giddings:2001bu}
S.~B.~Giddings and S.~Thomas,
``High energy colliders as black hole factories: The end of short  distance physics,''
Phys.\ Rev.\ D {\bf 65}, 056010 (2002)


\bibitem{Dimopoulos:2001hw}
S.~Dimopoulos and G.~Landsberg,
``Black holes at the LHC,''
Phys.\ Rev.\ Lett.\  {\bf 87}, 161602 (2001)

\bibitem{Cavaglia:2002si}
M.~Cavaglia,
``Black hole and brane production in TeV gravity: A review,''
hep-ph/0210296.


\bibitem{Randall:1999ee}
L.~Randall and R.~Sundrum,
 ``A large mass hierarchy from a small extra dimension,''
Phys.\ Rev.\ Lett.\  {\bf 83}, 3370 (1999)
 

\bibitem{Randall:1999vf}

L.~Randall and R.~Sundrum,
 ``An alternative to compactification,''
Phys.\ Rev.\ Lett.\  {\bf 83}, 4690 (1999)
 

\bibitem{Myers:un}
R.~C.~Myers and M.~J.~Perry,
``Black Holes In Higher Dimensional Space-Times,''
Annals Phys.\  {\bf 172}, 304 (1986).


\bibitem{Tangherlini}
F.R. Tangherlini, Nuovo Cimento {\bf 27}, 636 (1963).

\bibitem{Myers:rx}
R.~C.~Myers,
``Higher Dimensional Black Holes In Compactified Space-Times,''
Phys.\ Rev.\ D {\bf 35}, 455 (1987).


\bibitem{Bogojevic:1990hv}
A.~R.~Bogojevic and L.~Perivolaropoulos,
``Black Holes In A Periodic Universe,''
Mod.\ Phys.\ Lett.\ A {\bf 6}, 369 (1991).



\bibitem{Gregory:1993vy}
R.~Gregory and R.~Laflamme,
 ``Black strings and p-branes are unstable,''
Phys.\ Rev.\ Lett.\  {\bf 70}, 2837 (1993), 
R.~Gregory,
``Black string instabilities in anti-de Sitter space,''
Class.\ Quant.\ Grav.\  {\bf 17}, L125 (2000)
 


\bibitem{Hirayama:2001bi}
T.~Hirayama and G.~Kang,
``Stable black strings in anti-de Sitter space,''
Phys.\ Rev.\ D {\bf 64}, 064010 (2001)
 


\bibitem{Dadhich:2000am}
N.~Dadhich, R.~Maartens, P.~Papadopoulos and V.~Rezania,
 ``Black holes on the brane,''
Phys.\ Lett.\ B {\bf 487}, 1 (2000)
 
 
\bibitem{Giannakis:2001ss}
I.~Giannakis and H.~C.~Ren,
 ``Possible extensions of the 4-D Schwarzschild horizon in the brane  world,''
Phys.\ Rev.\ D {\bf 63}, 125017 (2001)
  
 
\bibitem{Chamblin:2001ra}
A.~Chamblin, H.~S.~Reall, H.~A.~Shinkai and T.~Shiromizu,
 ``Charged brane-world black holes,''
Phys.\ Rev.\ D {\bf 63}, 064015 (2001)
 

   \bibitem{Cadeau:2001tj}
C.~Cadeau and E.~Woolgar,
 ``New five dimensional black holes classified by horizon geometry, and a  Bianchi VI braneworld,''
Class.\ Quant.\ Grav.\  {\bf 18}, 527 (2001)
 

 
\bibitem{Vacaru:2001rf}
S.~I.~Vacaru and E.~Gaburov,
 ``Anisotropic black holes in Einstein and brane gravity,''
 hep-th/0108065.

  
\bibitem{Vacaru:2001wc}
S.~I.~Vacaru,
 ``A new method of constructing black hole solutions in Einstein and 5D  gravity,''
 hep-th/0110250.
 
 
\bibitem{Kanti:2001cj}
P.~Kanti and K.~Tamvakis,
 ``Quest for localized 4-D black holes in brane worlds,''
Phys.\ Rev.\ D {\bf 65}, 084010 (2002)

 


\bibitem{Casadio:2001jg}
R.~Casadio, A.~Fabbri and L.~Mazzacurati,
``New black holes in the brane-world?,''
Phys.\ Rev.\ D {\bf 65}, 084040 (2002)
 
 
\bibitem{Casadio:2002uv}
R.~Casadio and L.~Mazzacurati,
``Bulk shape of brane-world black holes,''
gr-qc/0205129. 


\bibitem{Kofinas:2002gq}
G.~Kofinas, E.~Papantonopoulos and V.~Zamarias,
``Black hole solutions in braneworlds with induced gravity,''
hep-th/0208207. 
 
 
\bibitem{Sengupta:2002fr}
G.~Sengupta,
``Rotating black holes in higher dimensional brane worlds,''
hep-th/0205087.
 
 
\bibitem{Modgil:2001hm}
M.~S.~Modgil, S.~Panda and G.~Sengupta,
``Rotating brane world black holes,''
Mod.\ Phys.\ Lett.\ A {\bf 17}, 1479 (2002)
 
 
\bibitem{Kodama:2002kj}
H.~Kodama,
``Vacuum branes in D-dimensional static spacetimes with spatial symmetry  IO(D-2), O(D-1) or O+(D-2,1),''
gr-qc/0204042.
 

\bibitem{Chamblin:2000by}
A.~Chamblin, S.~W.~Hawking and H.~S.~Reall,
 ``Brane-world black holes,''
Phys.\ Rev.\ D {\bf 61}, 065007 (2000).

 
\bibitem{Emparan:2001wk}
R.~Emparan and H.~S.~Reall,
``Generalized Weyl solutions,''
Phys.\ Rev.\ D {\bf 65}, 084025 (2002)


\bibitem{Emparan:2001wn}
R.~Emparan and H.~S.~Reall,
 ``A rotating black ring in five dimensions,''
Phys.\ Rev.\ Lett.\  {\bf 88}, 101101 (2002)


\bibitem{Shiromizu:2000wj}
T.~Shiromizu, K. I. ~Maeda and M.~Sasaki,
 ``The Einstein equations on the 3-brane world,''
Phys.\ Rev.\ D {\bf 62}, 024012 (2000);
M.~Sasaki, T.~Shiromizu and K. I. ~Maeda,
 ``Gravity, stability and energy conservation on the Randall-Sundrum  brane-world,''
Phys.\ Rev.\ D {\bf 62}, 024008 (2000)


\bibitem{Garriga:2000yh}
J.~Garriga and T.~Tanaka,
 ``Gravity in the brane-world,''
Phys.\ Rev.\ Lett.\  {\bf 84}, 2778 (2000)

\bibitem{Tanaka:2000er}
T.~Tanaka and X.~Montes,
 ``Gravity in the brane-world for two-branes model with stabilized  modulus,''
Nucl.\ Phys.\ B {\bf 582}, 259 (2000)

\bibitem{Mukohyama:2001ks}
S.~Mukohyama and L.~Kofman,
 ``Brane gravity at low energy,''
Phys.\ Rev.\ D {\bf 65}, 124025 (2002)
 
\bibitem{Tanaka:2000zv}
T.~Tanaka,
 ``Asymptotic behavior of perturbations in Randall-Sundrum brane-world,''
Prog.\ Theor.\ Phys.\  {\bf 104}, 545 (2000)

\bibitem{Giannakis:2001zx}
I.~Giannakis and H.~ C.~Ren,
 ``Recovery of the Schwarzschild metric in theories with localized gravity  beyond linear order,''
Phys.\ Rev.\ D {\bf 63}, 024001 (2001)

\bibitem{Kudoh:2001wb}
H.~Kudoh and T.~Tanaka,
 ``Second order perturbations in the Randall-Sundrum infinite brane-world  model,''
Phys.\ Rev.\ D {\bf 64}, 084022 (2001) 
 
\bibitem{Kudoh:2001kz}
H.~Kudoh and T.~Tanaka,
 ``Second order perturbations in the radius stabilized Randall-Sundrum two  branes model,''
Phys.\ Rev.\ D {\bf 65}, 104034 (2002)
 
\bibitem{Kudoh:2002mn}
H.~Kudoh and T.~Tanaka,
 ``Second order perturbations in the radius stabilized Randall-Sundrum two  branes model. II: Effect of relaxing strong coupling approximation,''
hep-th/0205041.


  
\bibitem{Emparan:2000fn}
For example, 
R.~Emparan, R.~Gregory and C.~Santos,
``Black holes on thick branes,''
Phys.\ Rev.\ D {\bf 63}, 104022 (2001)
Y.~Morisawa, R.~Yamazaki, D.~Ida, A.~Ishibashi and K.~i.~Nakao,
 ``Thick domain walls intersecting a black hole,''
Phys.\ Rev.\ D {\bf 62}, 084022 (2000)


\bibitem{Emparan:2000wa}
R.~Emparan, G.~T.~Horowitz and R.~C.~Myers,
 ``Exact description of black holes on branes,''
JHEP{\bf 0001}, 007 (2000);
``Exact description of black holes on branes. II: Comparison with BTZ  black holes and black strings,''
JHEP {\bf 0001}, 021 (2000)


\bibitem{Horowitz:2001cz}
G.~T.~Horowitz and K.~Maeda,
 ``Fate of the black string instability,''
Phys.\ Rev.\ Lett.\  {\bf 87}, 131301 (2001)

  
\bibitem{Wiseman:2002zc}
T.~Wiseman,
 ``Static axisymmetric vacuum solutions and non-uniform black strings,''
hep-th/0209051;
T.~Wiseman,
``From black strings to black holes,'' hep-th/0211028.


\bibitem{Tanaka:2002rb}
T.~Tanaka,
``Effective Gravity in Randall-Sundrum Infinite Brane World,''
Int. J. Theor. Phys. {\bf 41}, 2287-2297 (2002); 
T.~Tanaka,
 ``Classical black hole evaporation in Randall-Sundrum infinite  braneworld,''
 gr-qc/0203082. 
 
 
\bibitem{Emparan:2002px}
R.~Emparan, A.~Fabbri and N.~Kaloper,
``Quantum black holes as holograms in AdS braneworlds,''
JHEP {\bf 0208}, 043 (2002)
 
 
\bibitem{Wiseman:2001xt}
T.~Wiseman,
 ``Relativistic stars in Randall-Sundrum gravity,''
Phys.\ Rev.\ D {\bf 65}, 124007 (2002)
 
\bibitem{Nakamura:1981kc}
T.~Nakamura, K.~i.~Maeda, S.~Miyama and M.~Sasaki,
``General Relativistic Collapse of an Axially Symmetric Star. I, ''
Prog.\ Theor.\ Phys.\  {\bf 63}, 1229 (1980).

 

\bibitem{Eardley:1997hk}
See for example, D.~M.~Eardley,
 ``Black hole boundary conditions and coordinate conditions,''
Phys.\ Rev.\ D {\bf 57}, 2299 (1998)



\bibitem{Wald:1984}
R.~M.~Wald, 
\textit{General Relativity}
(The University of Chicago Press, 1984)

\bibitem{Choptuik:2003}
M.W. Choptuik, E. W. Hirschmann, S. L. Liebling, and F. Pretorius,
``An axisymmetric gravitational collapse code,"
gr-qc/0301006.~~
M. Alcubierre, S. Brandt, B. Bruegmann, D. Holz, E. Seidel, R. Takahashi and J. Thornburg
``Symmetry without Symmetry: Numerical Simulation of Axisymmetric Systems using Cartesian Grids"
Int.\ J.\ Mod.\ Phys. D10 273-290 (2001).
  
\bibitem{Gibbons:2002ju}
G.~W.~Gibbons, D.~Ida and T.~Shiromizu,
``Uniqueness of (dilatonic) charged black holes and black p-branes in  higher dimensions,''
Phys.\ Rev.\ D {\bf 66}, 044010 (2002)
 
 
\bibitem{Gibbons:2002av}
G.~W.~Gibbons, D.~Ida and T.~Shiromizu,
``Uniqueness and non-uniqueness of static black holes in higher  dimensions,''
Phys.\ Rev.\ Lett.\  {\bf 89}, 041101 (2002)

\end{thebibliography}
\end{document}